\documentclass[usenatbib,useAMS]{mn2e}
\usepackage{graphics,time}
\usepackage{amsmath,txfonts}
\usepackage{array,color,amssymb,enumerate}
\def\del#1{{}}

\bibliography{plain}
\input{epsf}
\newcommand{\be}{\begin{equation}}
\newcommand{\ee}{\end{equation}}
\newcommand{\ba}{\begin{eqnarray}}
\newcommand{\ea}{\end{eqnarray}}
\newcommand{\no}{\nonumber}
\newcommand{\bi}{\begin{itemize}}
\newcommand{\ei}{\end{itemize}}
\newcommand{\hMpc}{$h\,$Mpc$^{-1} $}
\newcommand{\Mpch}{$h^{-1}\,$Mpc }
\newcommand{\bfi}{\begin{figure}
\epsfxsize=9cm
\epsffile}
\newcommand{\efi}{\end{figure}}
\newcommand{\mnras}{MNRAS}
\newcommand{\apj}{ApJ}
\newcommand{\apjl}{ApJ}

\newcommand{\prd}{PRD}
\newcommand{\physrep}{Physics Reports}
\newcommand{\araa}{Annual Review of Astronomy \& Astrophysics}

\title{The kinetic SZ tomography with spectroscopic redshift surveys}
\author[Shao et al.]{Jiawei Shao$^{1,2}$\thanks{Email:jwshao@shao.ac.cn},
Pengjie Zhang$^1$\thanks{Email:pjzhang@shao.ac.cn}, Weipeng Lin$^1$, Yipeng
Jing$^1$, Jun Pan$^3$
\\$^1${Key Laboratory for Research in Galaxies and
Cosmology, Shanghai Astronomical Observatory, Nandan Road 80, Shanghai,
200030, China} 
\\$^2${Graduate School of the Chinese Academy of Sciences, 19A,
Yuquan Road, Beijing, China}
\\$^3$The Purple Mountain Observatory, 2 West Beijing Road, Nanjing 210008,
China}
\begin{document}
\maketitle
\begin{abstract}
The kinetic Sunyaev Zel'dovich effect (kSZ) effect is a potentially powerful probe to
the missing baryons. However, the kSZ signal is overwhelmed by various
contaminations and the cosmological application is hampered by loss of
redshift information due to the projection effect.  We propose a kSZ
tomography method to alleviate these problems, with the aid of galaxy
spectroscopic redshift surveys. We propose to estimate the large scale
peculiar velocity through the 3D galaxy distribution, weigh it by the 3D
galaxy density and adopt the product projected along the line of sight with a
proper weighting as an estimator of the true  kSZ temperature fluctuation
$\Theta$. Since the  underlying directional dependence in the estimator
$\hat{\Theta}$ closely resembles that in the true kSZ signal $\Theta$,
$\hat{\Theta}$ is tightly correlated with $\Theta$. It thus avoids the problem
of null correlation between the galaxy density and $\Theta$, which prohibits
the kSZ extraction through the usual density-CMB two-point cross correlation
measurement. We thus propose to measure the kSZ signal through the
$\Hat{\Theta}$-$\Theta$ cross correlation.  This approach has a number of
advantages. (1) Due to the underlying directional dependence of
$\hat{\Theta}$, it is  uncorrelated with the primary CMB, the  thermal SZ
effect and astrophysical contaminations such as the dusty star forming
galaxies.  Thus the $\hat{\Theta}$-$\Theta$ cross correlation picks up the kSZ
signal in the SZ survey with a clean manner. (2) With the aid of galaxy
redshifts, the cross correlation recovers the redshift information of the kSZ
signal and allows for more detailed investigation on missing baryons.  (3)
Since the galaxy surveys usually have high S/N,  the S/N of the kSZ
measurement through the $\hat{\Theta}$-$\Theta$ cross correlation can be
significantly improved. 

We test the proposed kSZ tomography against non-adiabatic and adiabatic
hydrodynamical simulations. We confirm that $\hat{\Theta}$ is indeed tightly
correlated with $\Theta$ at  $k\la 1h/$Mpc,  although nonlinearities in the
density and velocity fields and  nonlinear redshift distortion  do weaken the
tightness of the $\hat{\Theta}$-$\Theta$ correlation.  We further quantify the
reconstruction noise in $\Hat{\Theta}$ from galaxy distribution shot noise.
Based on these results, we quantify the  applicability of the proposed kSZ
tomography for future surveys. We find that, in combination with the BigBOSS-N
spectroscopic redshift survey, the PLANCK CMB experiment will be able to 
detect the kSZ with an overall significance of $\sim 50\sigma$ and further
measure its redshift distribution  at many redshift bins over $0<z<2$. 
\end{abstract}

\section{Introduction}
Robust evidences from CMB and BBN show that the baryonic matter accounts for
$\sim 4\%$ of the total matter and energy of the universe. However, only a
fraction of this baryon budget has been detected in the local universe, either
in the form of stars, interstellar medium (ISM) and intracluster medium (ICM)
(refer to \citealt{Fukugita2004, Fukugita2006} for census of
the baryon budget), while $\sim 50\%$ of the baryons remains elusive to
robust direct detection.

Looking for these ``missing'' baryons is crucial for the validity of our
standard cosmology model. The standard theory of hierarchical structure
formation models indicates that the majority of baryons exist between galaxies
as the diffuse intergalactic medium (IGM). Numerical simulations in the standard
cosmology further suggest that a large portion of this IGM is in the form of
warm-hot intergalactic medium (WHIM) \citep{CO99,Dave2001,CO06} with
temperature $10^5$K $<$ T $<10^7$ K, which is believed to reside in moderately
overdense structures such as filaments. On one hand, this kind intergalactic
medium is ionized dominantly by collisions, and is transparent to Ly$\alpha$
radiation, and thus is hard to trace by Ly$\alpha$ forest. The absorption
signature toward a bright X-ray source is also too weak to be resolved by
current spectrographs. On the other hand, though WHIM emits radiation in the
UV and soft X-ray bands, the emission strength is too weak for current
instruments to detect. There have been observations trying to detect WHIM via
absorption and radiation signature, but many of them are of low significance
or misinterpreted. See \citep{Bregman2007} for a recent review of the
status of WHIM detections.

Among the methods of probing missing baryons, the Sunyaev Zel'dovich (SZ)
effect is one of the most promising. Our universe is almost completely
ionized after $z=6$, where free electrons are prevailing in galaxy clusters as
high energy ICM and in the less overdense filamentary structures as IGM. Free
electrons will scatter off CMB photons through inverse Compton scattering, and
generate secondary CMB temperature anisotropies, which is known as the Sunyaev
Zel'dovich effect.  Therefore, the SZ effect is contributed by virtually all
electrons. In principle, from the SZ observations, we are able to find all
electrons and hence all baryons, due to the electric neutrality of the
universe.

The two major categories of the SZ effect are the thermal SZ effect (tSZ), arising
through the thermal motion of electrons, and the kinetic SZ effect (kSZ),
arising through the bulk motion of electrons. The efficiency for given
electrons to generate the SZ effect is proportional to the thermal temperature for
the thermal SZ effect and is proportional to the bulk peculiar velocity for
the kinetic SZ effect. Since the thermal temperature of electrons strongly
couples to the electron density, the dominant contribution to the thermal SZ
effect comes from the high-density and high-temperature ICM. For example,
\citet{White2002} found that 75\% of the total thermal SZ effect at multipole
$\ell<2000$ comes from virialized regions with gas overdensity $\delta_{\rm
gas}>100$ and \citet{H-Monteagudo2006} found similar results: $80\%$ of the
tSZ signal comes from collapsed structures.  Since the missing baryons are
likely in less dense regions with lower temperature,  the ability to find
missing baryons through the  thermal SZ effect is limited.

On the other hand, the kinetic SZ effect has a better potential to probe the
missing baryons. The peculiar velocity is determined by the large scale
gravitational potential and is thus only weakly coupled to local mass
concentration. Hence, the contribution to the kSZ effect is roughly proportional
to the total mass of each baryon component, e.g., ICM and IGM. Namely, it is an
approximately unbiased probe of baryons, regardless of their thermal state (as
long as they are ionized). Since the mass fraction of the missing baryons is
$\sim 50\%$, we thus expect a comparable contribution to the kSZ effect from the
missing baryons. This makes the kinetic SZ effect a promising probe of the
missing baryons. For example, there have been works studying the
contribution of kSZ effect by WHIM to the CMB anisotropies
\citep{A-Barandela2008, Santos2009}.

The remaining question is to measure the kinetic SZ effect, for which there
are several obstacles. First of all, the signal of the kinetic SZ effect is
not only weak, but also lack of spectral features to extract from the
overwhelming primary CMB. Thus, from the CMB measurement alone, we only expect
to detect it in the auto power spectrum measurement at $\ell\ga 3000$, where
the primary CMB damps significantly and the kinetic SZ effect begins to
dominate (e.g. \citealt{Ma2002,Zhang2004}), and at $\nu=217$ GHz frequency
band, where the (non-relativistic) thermal SZ effect vanishes. However,
refer to an interesting paper by \citealt{Nozawa2006} concerning the residual
tSZ effect at $217$GHz due to the relativistic corrections by massive
clusters. Other contaminations such as the dusty star-forming galaxies
\citep{Hall2010} make the kSZ detection even more difficult.  For these reasons,
even for CMB experiment as advanced as Planck, which has multiple bands over
wide frequency range, it is difficult to detect the kinetic SZ effect directly,
given its limited angular resolution. Secondly, like the thermal SZ effect, the
kinetic SZ effect suffers from severe projection effect. It measures the
electron momentum projected along the line of sight, thus the redshift
information of baryons is entangled in the projection with a projection length
of the size of the horizon. This not only degrades its power to infer the
evolution of missing baryons over the cosmic epoch, but could also lead to large
bias. For example, the patchy reionization could contribute a significant
fraction to the kinetic SZ effect \citep{Zahn2005, McQuinn2005, Iliev2007}.
Without redshift information, unless the reionization process is well understood
and interpreted appropriately, the extra contribution from patchy reionization
could be mis-interpreted as the sign of missing baryons. It's interesting
to notice the work of \citet{H-Monteagudo2009}, which proposed to recover the
signature of the bulk flow of the missing baryons by cross correlating future
CMB data sets with kSZ estimates in galaxy clusters.

In this paper, we propose a kinetic SZ tomography method to overcome the above
obstacles. The basic idea is to cross correlate the CMB observation with a
galaxy redshift survey or other surveys of the large scale structure with
sufficiently accurate redshift information such as the 21cm intensity mapping
\citep{Chang08}. Due to the cancellation mechanism arising from the underlying
directional dependence of the kinetic SZ signal as a result of its
vector nature, direct cross correlation between the kinetic SZ effect and the
galaxy number density vanishes (refer to Fig. \ref{fig:cancel} for more
detailed explanation). This differs significantly from the tight correlation
between the thermal SZ effect and the galaxy density \citep{ZP2001, Shao2009}.
One way to circumvent this cancellation is to square the kinetic SZ effect and
measure its correlation with galaxies \citep{Dore2004,DeDeo2005}. 

We propose an alternative approach to avoid the cancellation. One can
reconstruct the peculiar velocity field from the galaxy spectroscopic redshift
survey, weigh it with the observed galaxy number density and other
redshift-dependent factors to reconstruct a weighted momentum field. This
weighted {\it momentum} field has roughly the same directional dependence as
the true kSZ signal and thus avoids the cancellation. We thus expect a
measurable cross correlation between the reconstructed field and the CMB map,
which extracts the kSZ component in the CMB temperature fluctuation.

The cross correlation signal should arise solely from the corresponding redshift
range where galaxies reside, to an excellent approximation for a reasonably
thick galaxy redshift bin ($\Delta z\ga 0.2$) and sufficiently small angular
scales ($\ell\ga 10$). Thus it disentangles the kinetic SZ contribution
in this redshift range from contributions of any other redshifts and recovers
the redshift information of the kinetic SZ effect. It is for this reason that we
dubbed this method as {\it the kinetic SZ tomography}, analogous to the thermal SZ
tomography and also to the well known lensing tomography. An immediate
application of the kinetic SZ tomography is to separate the patch reionization
from the late time kinetic SZ effect. It is also effective to eliminate
contaminations from the primary CMB, thermal SZ and other foreground
contaminations, which do not have the characteristic directional
dependence and thus should be uncorrelated with the reconstructed momentum
field. Since the galaxy surveys usually have excellent signal to noise (S/N),
the cross correlation measurement can achieve much higher S/N than the auto
correlation measurement of the kinetic SZ effect from the SZ surveys alone. 
Later we will show that, PLANCK plus BigBOSS can detect the kinetic SZ effect at
$\sim 50\sigma$ level. 

\bfi{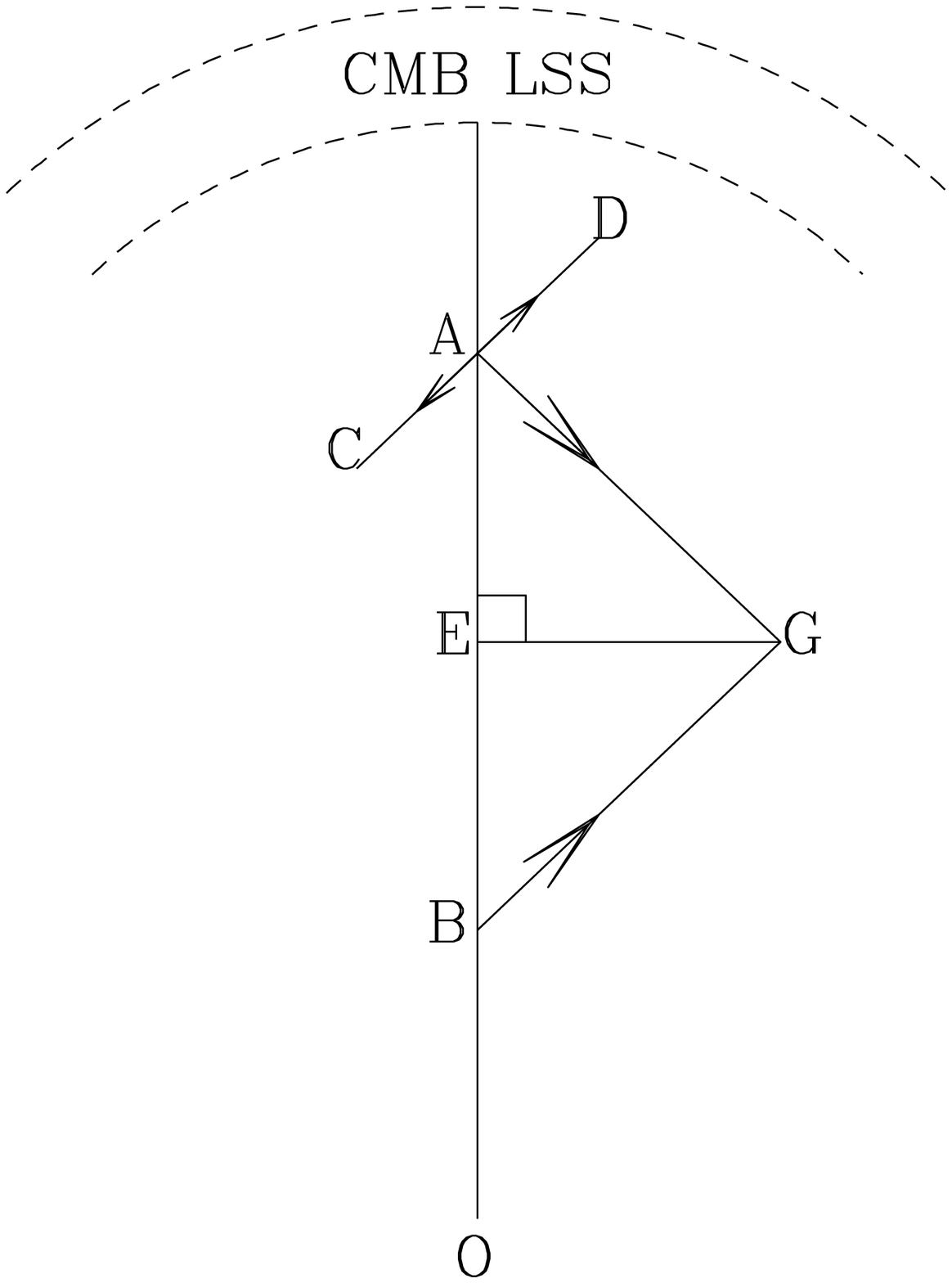}
\caption{The vanishing correlation between the kinetic SZ effect and the
galaxy number density. The line of sight is from the observer (O), through the
points B, E and A,  to the last scattering surface. One galaxy resides at
point G and we shall discuss the velocity distribution given the existence of
this galaxy.  The line EG is perpendicular to the line of sight and AE=EB. The
velocity at any point along the line of sight, e.g., the point A, can be
decomposed, with respect to the position of the given galaxy, into one
component along the direction AG and one perpendicular to the direction AG.
The perpendicular component has equal probability to be along the direction AC
or the direction AD and thus the net contribution to the kinetic SZ-galaxy
cross correlation is zero. So the only velocity component which may contribute
is the one  along the direction AG. However, from the symmetry argument, if
the evolution effect (light cone effect) can be neglected, velocity at point B
has an equal probability to have a component along the direction BG, with the
same amplitude. The projection of the two onto the line of sight cancels
exactly. So the cross correlation between the kinetic SZ effect and the galaxy
density vanishes.  \label{fig:cancel}} 
\efi

At the beginning stage of this work, \citet{Ho2009} published a work based on
similar idea, whose applicability is further confirmed in this paper. The two
works are carried out independently and thus  differ in many details.  We test
the proposed kSZ tomography against a controlled set of hydrodynamical
simulations and quantify the tightness of correlations between the
reconstructed map and the true kSZ signal at various redshifts over $0<z<2$.
We further investigate several complexities such as redshift distortion and
feedback, and demonstrate the robustness of the kSZ tomography against various
complexities. Both works  confirm the power of the kSZ tomography. 

The paper is organized as follows. In \S \ref{sec:formalism}, we introduce the
kSZ tomography method and discuss its limitations in general. We then test it
against our hydrodynamic simulations in \S \ref{sec:feasibility}.  The
reconstruction and the test are done in both the real space (\S
\ref{subsec:real_space}) and the redshift space (\S
\ref{subsec:redshift_space}).  In this section, we approximate galaxies as
dark matter particles. So it corresponds to idealized surveys of virtually
infinite galaxies such that shot noise in galaxy distribution is negligible.
Realistic surveys have much lower galaxy number density and thus shot noise in
the galaxy distribution induces non-negligible reconstruction noise. We
directly quantify it from our simulations with a proper scaling (\S
\ref{subsec:shotnoise}). We are then able to forecast its performance for
survey combinations like the Planck CMB experiment plus the BigBOSS
spectroscopic redshift survey (\S \ref{sec:err_fore}). We discuss and conclude
in \S \ref{sec:conclusion}. We present more technical details and further
discussions in the two appendices. 

\section{The kinetic SZ tomography}
\label{sec:formalism}
Bulk motions of free electrons induce secondary CMB anisotropies, namely the
kinetic Sunyaev Zel'dovich (kSZ) effect \citep{SZ1972,SZ1980}, with
temperature fluctuations 
\ba 
\label{eqn:ksz}
\Theta(\hat{n})&\equiv &\frac{\Delta T|_{\rm kSZ}}{T_{\rm CMB}}
    =\int \chi_e \bar{n}_e \sigma_T \frac{(1+\delta_e){\bf v}\cdot \hat{n}}{c}
    \exp[-\tau(z)] a{\rm d}\chi \nonumber\\
    &\equiv &\int {\bf p}_{\parallel} W_{\rm kSZ}(z)d\chi \no\\
    &\equiv &\sum_i\Theta_i\ ;\ \Theta_i=\int_{\chi_i-\Delta
\chi_i/2}^{\chi_i+\Delta \chi_i/2} {\bf p}_{\parallel} W_{\rm kSZ}(z)d\chi \ ,
\ea
where $\Theta_i$ is the contribution from electrons in the $i$-th redshift
bin, spanning the comoving coordinate range ${\chi_i-\Delta
  \chi_i/2}<\chi<{\chi_i+\Delta \chi_i/2}$. 
$\chi$ is the comoving radial coordinate, $c$ is the speed of light and $\tau$
is the Thompson optical depth. ${\bf p}\equiv (1+\delta_e){\bf v}$ is the
(normalized) electron momentum and the subscript ``$\parallel$'' denotes
the projection along the line of sight. $\delta_e$ is the electron number
overdensity, and ${\bf v}$ is the electron peculiar velocity. The weighting
function $W_{\rm kSZ}=\chi_e\bar{n}_e\sigma_T\exp(-\tau)a/c$ modulates the
contribution from each redshift. Throughout the paper, we focus on the kSZ
effect after reionization and thus set $\chi_e=1$, an excellent approximation.
Due to the neutrality, $\delta_e=\delta_{\rm gas}$, where
$\delta_{\rm gas}$ is the overdensity of (ionized) gas. 

As a vector, ${\bf p}$ can be always decomposed into a gradient (irrotational)
part ${\bf p}_E$ and  curl (rotational) part ${\bf p}_B$,  ${\bf p}\equiv{\bf
  p}_E+{\bf p}_B$, where $\nabla \times {\bf p}_E=0$ and  $\nabla \cdot{\bf
  p}_B=0$ respectively. Here, the ``E'' and ``B'' notations are analogous to
the electromagnetic fields. As can be inferred from Eq. \ref{eqn:ksz}, the
contribution from the gradient part is largely canceled out when integrating
along the line of sight, as long as the weighting function $W_{\rm  kSZ}$ varies
slowly across a correlation length of ${\bf p}_E$, which is of the order $100$
Mpc$/h$ today.\footnote{This condition can be violated at the epoch of reionization,
  where the patchy reionization causes $W$ to vary significantly over $\sim
  10h/$Mpc scales. }. 
For the kinetic SZ effect after reionization, $W_{\rm  kSZ}$ only changes
significantly over the Hubble scale, thus the contribution from the gradient
part is negligible and the only significant contribution comes from  ${\bf
p}_B$ \citep{Vishniac1987}.

${\bf p}_B$ in general has two sources of contribution. Since $\nabla\times
{\bf p}_B=\nabla\times {\bf p}=(1+\delta_e)\nabla\times {\bf v}+\nabla
\delta_e\times {\bf v}$, the B-mode of ${\bf p}$ can come from the B-mode of
${\bf v}$ or from the cross talk between the density and velocity. Following
the same notation, we can decompose the velocity into a ``E'' mode (gradient
part) ${\bf v}_E$ and a ``B'' mode (rotational part) ${\bf v}_B$. For purely
gravitational interaction, the velocity ``B'' mode decays, until multi-streaming
and shell crossing arise due to the nonlinear evolution (see for example chapter
2 of \citealt{Bernardeau2002} for a discussion). Thus in the linear and weakly
nonlinear regimes a
good approximation is ${\bf v}={\bf v}_E$ and  the only contribution to the kSZ
effect comes from the cross talk between the density gradient and the velocity.
This is the well known Ostriker-Vishniac (OV) effect \citep{OV1986,
Vishniac1987}. In the nonlinear regime, ${\bf v}_B$ grows and can also contribute
to the kSZ effect \citep{Zhang2004}.

The kinetic SZ tomography that we propose requires  combination of
a SZ survey and a galaxy spectroscopic redshift survey with overlapping sky
coverage. It contains three major steps: 
\bi
\item Construct a 2D map $\hat{\Theta}_i$ from the 3D distribution of galaxies
  in the $i$-th redshift bin. Ideally, $\hat{\Theta}_i$ should be  tightly
  correlated with $\Theta_i$,  the true kinetic SZ signal from this redshift bin.  A crucial
  ingredient to guarantee a tight correlation is to estimate the  peculiar 
  velocity through the 3D galaxy distribution and use it to construct
  $\hat{\Theta}_i$. Hereafter, we often neglect the subscript ``$i$'' where it
  does not cause confusion.   
\item Cross correlate the reconstructed $\hat{\Theta}$ with a overlapping SZ
  survey. The cross correlation signal, to an excellent approximation, solely
  comes from the kinetic SZ in the chosen redshift bin. It is this step
  that recovers the redshift information of the kinetic SZ effect, eliminates
  various contaminations and reduces statistical errors. 
\item Interpret the {\it measured} cross correlation signal and reconstruct
  the true kinetic SZ signal. Trick similar to the one adopted in the thermal
SZ tomography \citep{Shao2009} can also be applied here. 
\ei
The current paper will focus on the first two steps and only briefly discuss
the third step.

\subsection{The kinetic SZ reconstruction}
The primary goal of this paper is to extract the kSZ signal as well as its
redshift information through cross correlating SZ surveys with galaxy
spectroscopic redshift surveys. As explained early, the usual two-point cross
correlation between the kinetic SZ effect $\Theta$ and galaxy overdensity $\delta_g$
vanishes ($\langle \Theta \delta_g\rangle=0$), due to the cancellation of
positive and negative velocities along the line of sight (Fig.
\ref{fig:cancel}).  A natural step to avoid such cancellation is to recover
the velocity information and weigh the galaxies accordingly. This can be done
in spectroscopic galaxy redshift surveys. 

Spectroscopic galaxy redshift surveys, such as
LAMOST\footnote{http://www.lamost.org/website/en},
BOSS\footnote{http://cosmology.lbl.gov/BOSS/},
BigBOSS\footnote{http://bigboss.lbl.gov/index.html},
SKA\footnote{http://www.skatelescope.org/},
Euclid\footnote{http://sci.esa.int/euclid}
and JDEM/ADEPT\footnote{http://jdem.gsfc.nasa.gov/}, will measure the 3D galaxy
distribution $\delta_g({\bf x})$. At least part of them  will overlap with SZ
surveys such as ACT\footnote{http://www.physics.princeton.edu/act/index.html},
SPT\footnote{http://pole.uchicago.edu/} and
PLANCK\footnote{http://www.rssd.esa.int/index.php?project=planck} on the sky
coverage.  We are able to recover the velocity field from $\delta_g({\bf x})$ and
then $\hat{\Theta}$, the galaxy momentum properly weighted. 
Since the  direction of the peculiar velocity is taken into account in
$\hat{\Theta}$, the cross  correlation $\langle \hat{\Theta}\Theta\rangle$ no
longer suffers from the usual cancellation and thus $\langle
\hat{\Theta}\Theta\rangle\neq 0$.

In the linear regime, the mass conservation (i.e. {\it the continuity
equation}) reduces to
 \be
 \dot{\delta}_m+\nabla\cdot {\bf v}=0\ ,
 \label{eqn:massconservation1}
\ee 
where $\delta_m$ is the matter overdensity and ${\bf v}$ is the peculiar
velocity of the matter field. In the same regime, namely at large scale, the
3D galaxy distribution is a good proxy of the underlying 3D matter distribution.
This can be described by a galaxy bias $\delta_g=b_g\delta_m$. Under the above
condition, given the observed 3D galaxy distribution $\delta^{\rm obs}_g({\bf
x})$, we are able to obtain an estimator of the velocity field, which in
Fourier space reads 
\be
  \label{eqn:v}
  \hat{{\bf v}}({\bf k})=
  -ifH\delta^{\rm obs}_g({\bf k})\frac{{\bf k}}{k^2}\ ,
\ee 
where ${\bf k}$ is the 3D wave vector, $f\equiv d\ln D/d\ln a$ and $D$ is the
linear density growth rate.  The superscript ``hat''  in ${\bf v}$ ($\hat{\bf
v}$) and in other symbols (e.g. $\hat{\Theta}$)  denotes the reconstructed
quantity.  In reality, to suppress the noise and stabilize the reconstruction,
we often apply some filters to the density field, before applying Eq.
\ref{eqn:v}. So $\delta^{\rm obs}_g$ should be treated as  the smoothed
density field. 

Eq. \ref{eqn:v} is a {\it biased} estimator of the true peculiar velocity ${\bf
v}$. (1) The galaxy bias causes $\hat{\bf v}$ to be overestimated by a factor
$b_g$. (2) Redshift distortion causes the observed density to deviate from the
underlying matter density and thus biases the velocity reconstruction. (3)
Nonlinearities in the density evolution causes deviation from Eq.
\ref{eqn:massconservation1}. (4) In the nonlinear regime where shell crossing
and multi-streaming happen, velocity vorticity (rotational part, or curl part)
develops, which is completely missed by the estimator Eq. \ref{eqn:v}. 

The imperfectness of the velocity reconstruction is not as severe as it looks,
for  the kinetic SZ tomography. Later we will show that the performance of the
kSZ tomography is insensitive to deterministic errors in the velocity
reconstruction. Thus a deterministic galaxy density and velocity bias,
uncertainties in $f$ and $H$, linear redshift distortion (the Kaiser effect)
do not degrade the kSZ  tomography. However, stochastic errors from stochastic
galaxy bias, nonlinearities in the evolution of density and velocity do.
Later we will quantify their impacts and show them to be moderate at relevant
scales.

 With the 3D density field $\delta_g$ and thereby the recovered 3D velocity
  field $\hat{\bf v}$ in hand, we can then reconstruct a weighted 2D momentum map
\be
  \label{eqn:hatT}
  \hat{\Theta}\equiv \int d\chi \hat{{\bf p}}_{\parallel} \hat{W}(z)\ ,
\ee
where $\hat{{\bf p}}\equiv (1+\delta^{\rm obs}_g)\hat{{\bf v}}({\bf x})$, and
$\hat{{\bf v}}({\bf x})$ is the inverse Fourier transform of $\hat{{\bf v}}({\bf
 k})$. We choose the weighting function $\hat{W}(z)=W_{\rm kSZ}(z)$. The
integral in Eq. \ref{eqn:hatT} is over the corresponding redshift bin.  Again, we
can also apply some filters to the density field, before taking the
product in $\hat{\bf p}$. These filters are not necessary to be the same as
the ones for the  velocity reconstruction. Furthermore, the density measurement in the
product may not even be the same as the one used in the velocity measurement,
as correctly pointed by \cite{Ho2009}. 

Here we want to clarify a likely confusing point. As discussed before, the
kinetic SZ effect is mainly contributed by ${\bf p}_B$ instead of ${\bf p}_E$.
On the other hand, the reconstructed $\hat{\bf v}$ in Eq. \ref{eqn:v} is
actually ${\bf p}_E$, which becomes clear later in Eq.
\ref{eqn:massconservation}. It thus seems that the reconstruction misses the
dominant contribution to the kinetic SZ effect and thus should fail to work.
However, there is an extra factor $(1+\delta_g)$ in the estimator
$\hat{\Theta}$ (Eq. \ref{eqn:hatT}). Recall that, in the OV effect
\citep{OV1986, Vishniac1987}, it is the cross-talk between the density
gradient and the curl-free velocity that generates a curl component in ${\bf p}$.
Here, the cross-talk between the density and  ${\bf p}_E$ generates a B-mode
in the reconstructed $\hat{\bf p}$,  which is tightly correlated with the true
${\bf p}_B$ on relevant scales. This explains the reasonable performance of
the reconstruction technique and the kinetic SZ tomography.

\subsection{How to quantify the kSZ tomography performance?}
\label{subsec:performance}
The estimator $\hat{\Theta}$ is certainly imperfect. It can have both
systematical and random offsets with respect to $\Theta$.  These deviations 
should be both scale and redshift dependent.  These deviations can be
visualized by a $\Theta$-$\hat{\Theta}$ 
plot. Alternatively, it can be quantified by two parameters,  $r$ and
$b_{\hat{\Theta}}$. $r$ describes the tightness of the
$\hat{\Theta}$-$\Theta$ correlation\footnote{$r$ that we define differs from
  the one defined in   \citet{Ho2009}. First, their $r$ is for the velocity
  field instead of the  
momentum field, 3D instead of 2D. Second, their $r$ is not the cross correlation
coefficient, but is actually analogous to $r/b$ of the 3D velocity field, in our
notation. } and $b_{\hat{\Theta}}$ is the bias in $\hat{\Theta}$ with respect
to $\Theta$. Clearly,
both $r$ and $b_{\hat{\Theta}}$ depend on the redshift range of galaxies used
for reconstruction. We want to quantify the redshift dependence of $r$ and
$b$. Thus later in the analysis we will choose a projection length across
which  evolutions in $r$ and $b_{\Theta}$ are negligible. For such
projection, $r$ and $b_{\hat{\Theta}}$ are  functions of $z$ and the 2D wave vector
$k_{\perp}$, the inverse of the perpendicular spatial separation, 
\be
r(k_\perp,z)\equiv
\frac{P_{\hat{\Theta}
\Theta}(k_\perp,z)}{\sqrt{P_{\hat{\Theta}}(k_\perp,
z)P_{\Theta}(k_\perp,z)}}\ ,
\label{eqn:r}
\ee
and 
\be
b_{\hat{\Theta}}(k_\perp,z)\equiv
\sqrt{\frac{P_{\hat{\Theta}}(k_\perp,z)}{P_{\Theta}(k_\perp,z)}}\
, 
\label{eqn:b}
\ee
Later we will recognize ${\bf k}_\perp$ as the perpendicular component of the
usual 3D wave vector ${\bf k}$. The $P$s are the corresponding power spectra.  

There is no guarantee that $b_{\hat{\Theta}}$ is close to unity, even in the
linear regime. As explained earlier, the velocity estimation is
biased. Further, in the product $\hat{\Theta}\propto (1+\delta_g^{\rm obs})
\hat{\bf v}$, $\delta_g^{\rm obs}$ is also biased with respect to $\delta_e$.
Later in the Appendix \ref{sec:app1} we will 
show the complicated behavior of $b_{\hat{\Theta}}$.

However, large deviation of $b_{\hat{\Theta}}$ from unity does not necessarily 
mean poor performance of the kinetic SZ tomography, for three reasons. (1)
First, the signal-to-noise ratio (S/N) of the $\langle \Theta \hat\Theta
\rangle$ measurement, is solely determined by $r$.  The S/N of each mode
(Fourier or multipole mode) is \ba
\label{eqn:SN}
\left(\frac{S}{N}\right)^2&=&\frac{P^2_{\Theta\hat{\Theta}}}
 {P^2_{\Theta\hat{\Theta}}+(P_{\Theta}+P^{N}_{\Theta})
(P_{\hat{\Theta}}+P^{N}_{\hat{\Theta}})}
\no\\
&=&\frac{1}{1+r^{-2}\left(1+\frac{P^{N}_{\Theta}}{P_{\Theta}}
\right)
 \left(1+\frac{P^{N}_{\hat{\Theta}}}{P_{\hat{\Theta}}}\right)}\ ,
\ea
where $P^{N}_{\Theta}$ and $P^{N}_{\hat{\Theta}}$ are the corresponding noise
power spectra. We find that the $b_{\hat{\Theta}}$ dependence drops out in the error
estimation. A rescaling $\hat{\Theta}\rightarrow b_{\hat{\Theta}}
\hat{\Theta}$ leaves no effect on  the S/N of the cross correlation
measurement, since both $P^{N}_{\hat{\Theta}}, 
P_{\hat{\Theta}}\propto b_{\hat{\Theta}}^2$ and $r$ is unchanged under this
scaling. (2) $b_{\Theta}\neq 1$ definitely affects the expectation value of
the cross correlation. However,  since in the theoretical 
interpretation of the measured cross correlation, one can take the galaxy bias,
redshift distortion and possibly other complexities into account and thus
avoid systematical errors induced by $b_{\Theta}\neq 1$ reasonably well. (3)
Based on the same technique in the thermal SZ tomography \citep{Shao2009},  we
can combine the cross correlation measurement $P_{\Theta\hat{\Theta}}$ and the
auto correlation measurement $P_{\hat{\Theta}}$ to obtain
$P_{\Theta}=P^2_{\Theta\hat{\Theta}}/(r^2 P_{\Theta})$. This estimation relies
on no information of $b_{\hat{\Theta}}$ and thus avoids the bias problem. 

For these reasons, in the main text we will focus on $r$ to quantify the
performance of the kinetic SZ tomography and leave discussion on
$b_{\hat{\Theta}}$ in the Appendix \ref{sec:app1}. 

\subsection{Origins of the stochasticity $r\neq 1$ }
\label{subsec:stochasticity}
 A number of approximations made in the  reconstruction pipeline cause the
 stochasticity in the $\Theta$-$\hat{\Theta}$ relation ($r\neq 1$). (1) First
 of all, as an approximation to the exact mass conservation 
equation
\be
\label{eqn:massconservation}
\dot{\delta}_m+\nabla \cdot (1+\delta_m){\bf v}=0\ ,
\ee
the starting point Eq. \ref{eqn:massconservation1} only holds where
$\delta_m\ll 1$. (2) From Eq. \ref{eqn:massconservation1} to Eq. \ref{eqn:v}, we
have made assumptions of linear evolution ($\delta_m(k,z)\propto
D(z)\delta_m(k,z_i)$), deterministic bias between $\delta_g$ and $\delta_m$, 
and curl-free velocity. None of these approximations are exact in the nonlinear
regime. (3) Even under these assumptions, Eq. \ref{eqn:massconservation1} and
\ref{eqn:v} only hold in  real space. Namely, we have implicitly assumed that
the observed $\delta_g^{\rm obs}$ is the galaxy overdensity $\delta_g$ in real
space ($\delta_g^{\rm obs}=\delta_g$). However, in reality, what we directly
measure is the galaxy number overdensity $\delta_g^s$ in redshift space
($\delta_g^{\rm obs}=\delta^s_g$). Due to the redshift distortion,
$\delta_g^s\neq \delta_g$.  The $\delta_g^s$-$\delta_g$ relation is
stochastic, due to nonlinear mapping between redshift and real space,
nonlinearities in both the density and velocity fields (e.g.
\citealt{White2009}) and velocity vorticity \citep{Carlson2009}. (4) Even if
we have perfect E-mode velocity reconstruction, we still have no handle on the
B-mode velocity, which contributes the kinetic SZ effect.  (5) In Eq.
\ref{eqn:hatT}, we multiply the reconstructed velocity with $1+\delta_g^{\rm
obs}$ instead of $1+\delta_e$.  The possible stochasticity between $\delta_e$
and $\delta_g$ also increases the stochasticity between $\Theta$ and
$\hat{\Theta}$.

With the aid of our hydrodynamic simulation, we are able to quantify the 
combined influence of all these factors on $r$, except for the stochastic
galaxy bias.  \cite{Ho2009} used the halo occupation model and N-body 
simulations to produce galaxy mock catalog. This approach captures the
galaxy stochasticity and shows that the kSZ tomography is robust against
it. Our simulations have relatively small box size ($100h^{-1}$Mpc) and hence
do not  allow us to follow the same approach. So we leave this issue for
further  investigation.

\section{Testing against hydrodynamical simulations}
\label{sec:feasibility}
We test the kinetic SZ tomography against our hydrodynamical simulations. The
simulations are run with the  GADGET2 code \citep{Springel05} in a
$\Lambda$CDM cosmology with parameters: $\Lambda=0.732$, $\Omega_0=0.268$,
$\Omega_b=0.044$ $h=0.71$, $\sigma_8=0.85$. The box size of the simulation is
$L=100$\Mpch on each side, in which $512^3$ dark matter particles and $512^3$
gas particles are initially seeded (See more details about the simulation in
\citealt{Jing06, Lin2006}).  We have an non-adiabatic run, in which gas
particles are allowed to cool and condense into collisionless star particles,
along with which SN feedback is taken into account.  We also have an adiabatic
run with the same cosmological parameters and the same initial conditions.  We
will focus on the non-adiabatic run, since it has better capture on the
gastrophysics and hence better modeling of the kinetic SZ effect.  Unless
otherwise specified, all simulations are based on this simulation. We also
analyze the adiabatic run to better understanding the generality of the kSZ
tomography (\S \ref{subsec:gastrophysics}).   We choose in this work a few
representative redshifts to quantify the feasibility of the tomography
technique, which is mainly characterized by the quantity $r(k_\perp,z)$.

Our hydrodynamic simulations have direct information of the gas momentum
distribution and thus the kinetic SZ effect. However, the simulations don't
simulate galaxies. To proceed, we approximate galaxies as simulation dark matter
particles.  Since the galaxy stochasticity is likely sub-dominant
\citep{BP2009,Baldauf2010} , this approximation is reasonable to estimate $r$
between $\hat{\Theta}$ and $\Theta$, although it indeed over-estimates
it. As demonstrated by \citet{Ho2009}, the kSZ tomography is robust against
the galaxy stochasticity. So we will leave this issue elsewhere. Since the
number density of simulation  
particles  is much higher than 
that of galaxies in any realistic surveys, shot noise in the $\Hat{\Theta}$
reconstruction is negligible. To use the measured $r$ for forecasting, we have
to take the shot noise inevitable in realistic surveys into account. We leave
this issue until next section. 

We test the kinetic SZ tomography at several typical redshifts $z=0, 0.53,
1.02, 2.08$. We reconstruct the velocity and consequently the momentum field 
from the  dark matter distribution in the corresponding simulation outputs.
At these redshifts, we project the 3D momentum field over a single box size
(100\Mpch) to get the 2D momentum maps.  We then compare these maps with the
maps of underlying kinetic SZ signal, which is directly measured by
multiplying the gas density and velocity distribution in the corresponding
simulation outputs. Since the simulations have information of peculiar
velocity, we carry out the reconstruction and comparison in both real space
and redshift space.

\subsection{The reconstruction in real space}
\label{subsec:real_space}
\subsubsection{The reconstruction procedure}
\bfi{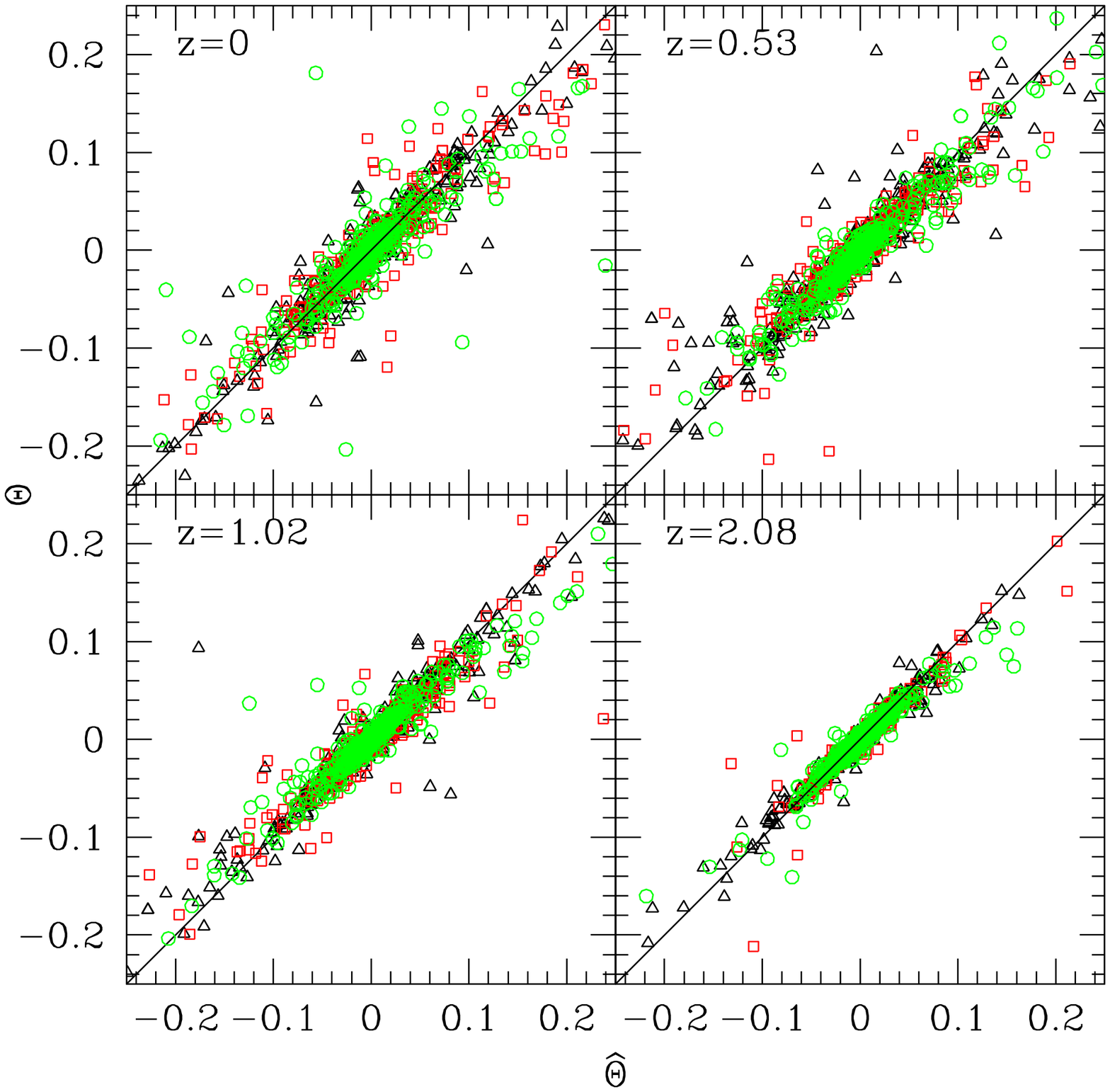}
\caption{The $\hat{\Theta}$-$\Theta$ map (in arbitrary unit) in the real
space at z=0, 0.53, 1.02, 2.08. In order to further reduce the scattering of
$\hat{\Theta}$-$\Theta$ relation, we average the recovered velocity field in
every neighboring $4^3=64$ cells to get the averaged velocity field $\hat{\bf
v}$ as well as the density field $\delta$ on a much coarser partitioning, say,
$64^3$ cells. We show here the cell-to-cell correspondence in all three
Cartesian directions by randomly selecting $512$ out of $64^3$ couples. The
recovered $\hat{\Theta}$ field intimately follows the kSZ signal along the
diagonal at high redshifts, while at low redshift, there are great fluctuations
due to non-linear evolution of the density field and thereby the velocity
field. The three components in Cartesian directions are denoted by different
point types, triangles for x, squares for y and circles for z. }
\label{fig:rmap64}
\efi

We first carry out the reconstruction in real space. The reconstruction is a
two-step process. The first step is to reconstruct the peculiar velocity. At
each redshift $z_i$, we construct the matter density field $\delta({\bf x})$
using clouds-in-cells (CIC) scheme in a partitioning of $n_g^3=256^3$ cells,
followed by a Fourier transform to get the matter distribution in Fourier space
$\delta({\bf k})$.  We then apply a Gaussian filter\footnote{\citet{Ho2009}
populated halos with galaxies. Since in that case the galaxy density is low, and
shot noise is non-negligible, they applied the Wiener filter to reduce the
Poisson noise. This is also a necessary procedure when dealing with real
data. In our work, we use simulation particles as galaxies, whose number
density is much higher and thus 
the shot noise is negligible. For this reason, we instead apply generally a
Gaussian window function to filter out small scale nonlinear fluctuations.} 
$W_{G}(k)=\exp({-k^2 R^2_s/2})$ to the original field $\delta({\bf k})$
in order to wipe off nonlinear fluctuations on small scales, which are
uncorrelated with the large scale velocity. The adopted smoothing length
$R_s=1.56$\Mpch applied in the whole context is around the radius of a typical
cluster, and we will discuss the influence of different smoothing length in \S
\ref{subsec:smoothing}. We then obtain the reconstructed velocity $\hat{\bf v}({\bf
k})$ from Eq. \ref{eqn:v} through this smoothed density field. By the inverse
Fourier transform, we obtain the real space velocity field $\hat{\bf v}({\bf
x})$. 

\bfi{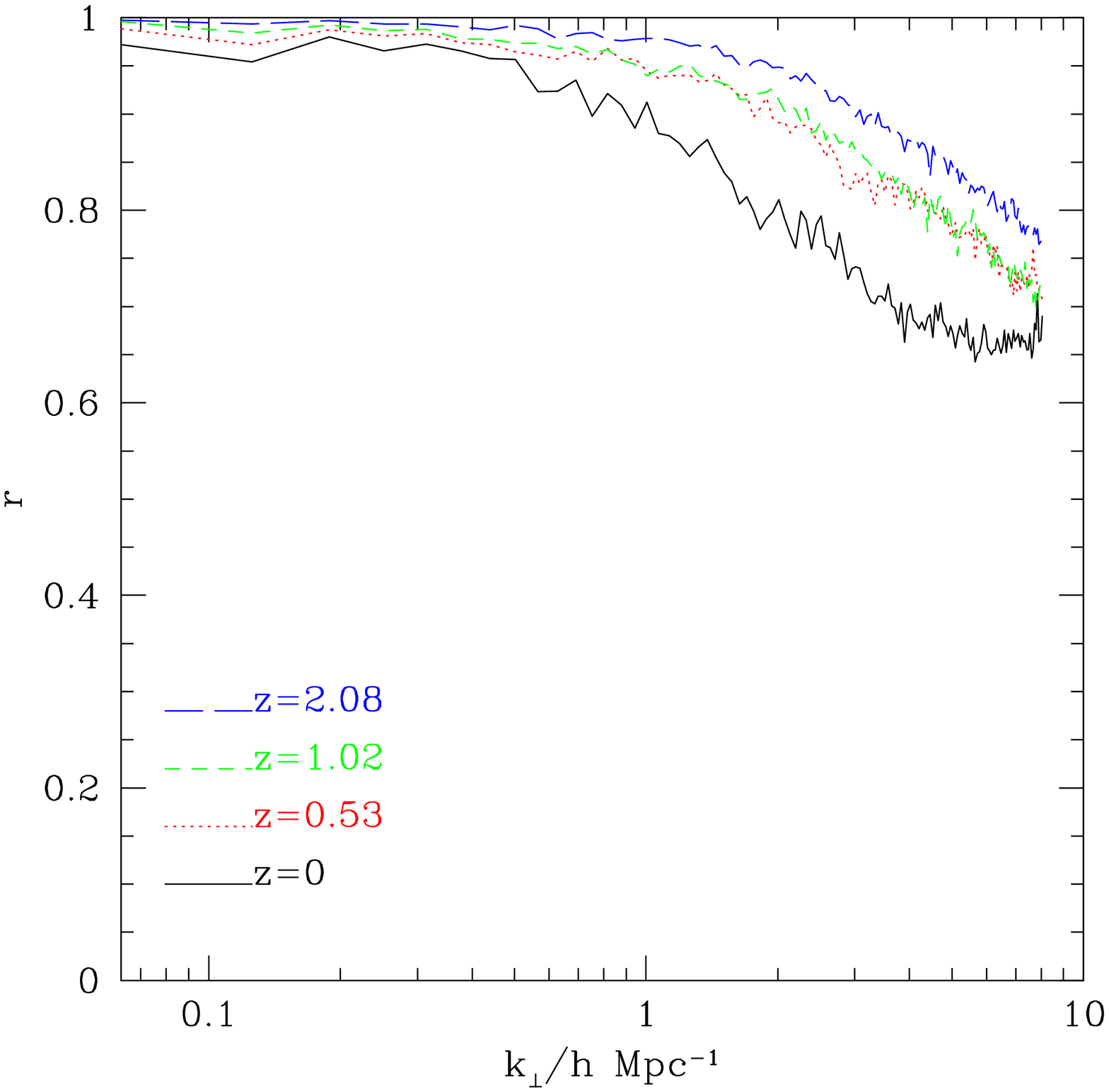}
\caption{The cross correlation coefficients $r$ between
$\hat{\Theta}$ and $\Theta$. The correlation coefficient is averaged over the
three components. The two are highly correlated, r$\simeq$0.9, on large scale
$k_\perp\le1$\hMpc for all the 4 redshifts, while less correlated to smaller
scales ends due to the non-linearities coming into play. Nevertheless, the
correlation coefficient is still tight at z=0, with $r>0.6$ up to
$k_{\perp}\sim 8$\hMpc.}
\label{fig:rcorr}
\efi

The second step is to project the momentum field along the line of sight
${\hat{n}}$ to get the reconstructed 2D map at redshift $z_i$
\begin{equation}
    \hat{\Theta}_{j1,j2}(\hat{\bf n})=\sum_{j3} {\hat{\bf n}} \cdot
\frac{\hat{\bf v}_{j1,j2,j3}}{c} (1+\delta_{j1,j2,j3})\ .
\label{eqn:RkSZ}
\end{equation} Since the weighting function $W_{\rm kSZ}$ changes little over
that scales, so we treat it as a constant and omit it. The sum is over a
single simulated box at the investigated redshift, instead of  stacking the
box in the light cone. Notice that the density term in the above equation (Eq.
\ref{eqn:RkSZ}) is the unsmoothed density,  which preserves the necessary
information of the true density field in the deriving of the momentum field.
In practice, we construct at each redshift three maps along three Cartesian
coordinate directions of the simulated box respectively. As they can be
considered as three independent measurements, in the following figures which
concern statistics, we show the average results unless specified. 

At the same time, we obtain the {\it true} kSZ signal in a similar way
\begin{equation}
    \Theta_{j1,j2}(\hat{\bf n})=\sum_{j3} \hat{\rm n}\cdot
\frac{{\bf v}_{b,j1,j2,j3}}{c}(1+\delta_{b,j1,j2,j3})\ ,
\end{equation}
where $\delta_b$ is the nonlinear overdensity of baryons. ${\bf v}_b$ is the
bulk velocity derived directly from the simulation by averaging within the
host cell (j1,j2,j3)
\be
    {\bf v}_b = \frac{\sum_i {\bf v}_{i} m_i w_i}{ \sum_i m_i w_i}\ ,
\ee
where ${\bf v}_i$ and $m_i$ are the comoving peculiar velocity and the mass of
$i$-th gas particle in the host cell. $w_i$ is a spline kernel used to smooth
the gas particles.

\subsubsection{The $\hat{\Theta}$-$\Theta$ relation}
In Figure \ref{fig:rmap64} we show the cell-to-cell $\hat{\Theta}$-$\Theta$
correspondence.   The $\hat{\Theta}$-$\Theta$ data points scatter around
$\hat{\Theta}=\Theta$, meaning a bias $b_{\Theta}\sim 1$. However, this should 
not be over-emphasized since this only represents an unrealistic case of galaxy bias
$b_g=1$ without redshift distortion. Reconstructions based on galaxy sample
with $b_g\neq 1$ would result in $b_{\Theta}\neq 1$ and hence a different slope of the
$\hat{\Theta}$-$\Theta$ relation.  Redshift distortion also changes the slope of the
$\hat{\Theta}^s$-$\Theta$ relation, as can be seen from Fig. \ref{fig:smap64}. 

The $\hat{\Theta}$-$\Theta$ relation shows non-negligible dispersion around
the mean, but still reasonably tight. This means that the stochasticity is
noticeable, but not yet overwhelming. We also notice that, the dispersion gets
stronger at lower redshift. This is not surprising, since the reconstruction is
based on linear theory and thus works better at higher redshift. Nonlinearities
in the density field and velocity field degrades the reconstruction accuracy, as
discussed in \S \ref{subsec:stochasticity}. 

\subsubsection{The cross correlation coefficient $r$}
\label{subsec:rcorr}
Out of the two quantities concerning to the reconstruction performance,
the cross correlation coefficient $r$ describes the tightness of
$\hat{\Theta}$-$\Theta$ relation and is a major measure of the kinetic SZ
tomography. $b_{\hat{\Theta}}$ is of less importance in quantifying the  kSZ
tomography performance, so we leave related results to the Appendix
\ref{sec:app1}. To measure them, we
perform 2D Fourier transforms of $\Theta$ ($\hat{\Theta}$).  $\Theta({\bf
k}_\perp)=\int d^2{\bf x}_{\perp} \Theta({\bf x}_{\perp}) \exp(i {\bf k}_\perp
\cdot{\bf x}_{\perp})/A$, where $A$ is the area of the map. Note here ${\bf
k_\perp}$ and ${\bf x}_{\perp}$ are both 2D variables. We then obtain the
power spectrum $P_{\Theta}$ with $(2\pi)^2 \delta_D({\bf k_\perp}-{\bf
k_\perp}^{\prime})P_\Theta (k_\perp)= \left<\Theta({\bf k_\perp})\Theta({\bf
k_\perp}^{\prime})\right>$. 

Fig. \ref{fig:rcorr} shows that the stochasticity in the
$\hat{\Theta}$-$\Theta$ relation is in general not a severe issue, even in the
strongly nonlinear regime, consistent with Fig. \ref{fig:rmap64}. As shown
we have strong correlations, $r\ga 0.9$ to $k_{\perp}=1$\hMpc at all
redshifts, and even $r\ga0.8$ to $k_{\perp}\sim3$\hMpc except z=0.
Nonlinearities do degrade the reconstruction, as we see that $r$ decreases
towards low redshifts. However, even at $z=0$, the $\hat{\Theta}$-$\Theta$
correlation is still pretty tight, with $r>0.6$ to $k_{\perp}\sim 8$\hMpc.
We don't correct for the aliasing effect, such that small scale results
may be misleading \citep{Jing2005}. Nevertheless, we do expect good results up to a
quarter of the Nyquist wavenumber, i.e. around $2$\hMpc, and it's safe for us
to estimate the cross power spectrum up to $\ell\sim2000-3000$.

\subsection{The reconstruction in redshift space}
\label{subsec:redshift_space}
\bfi{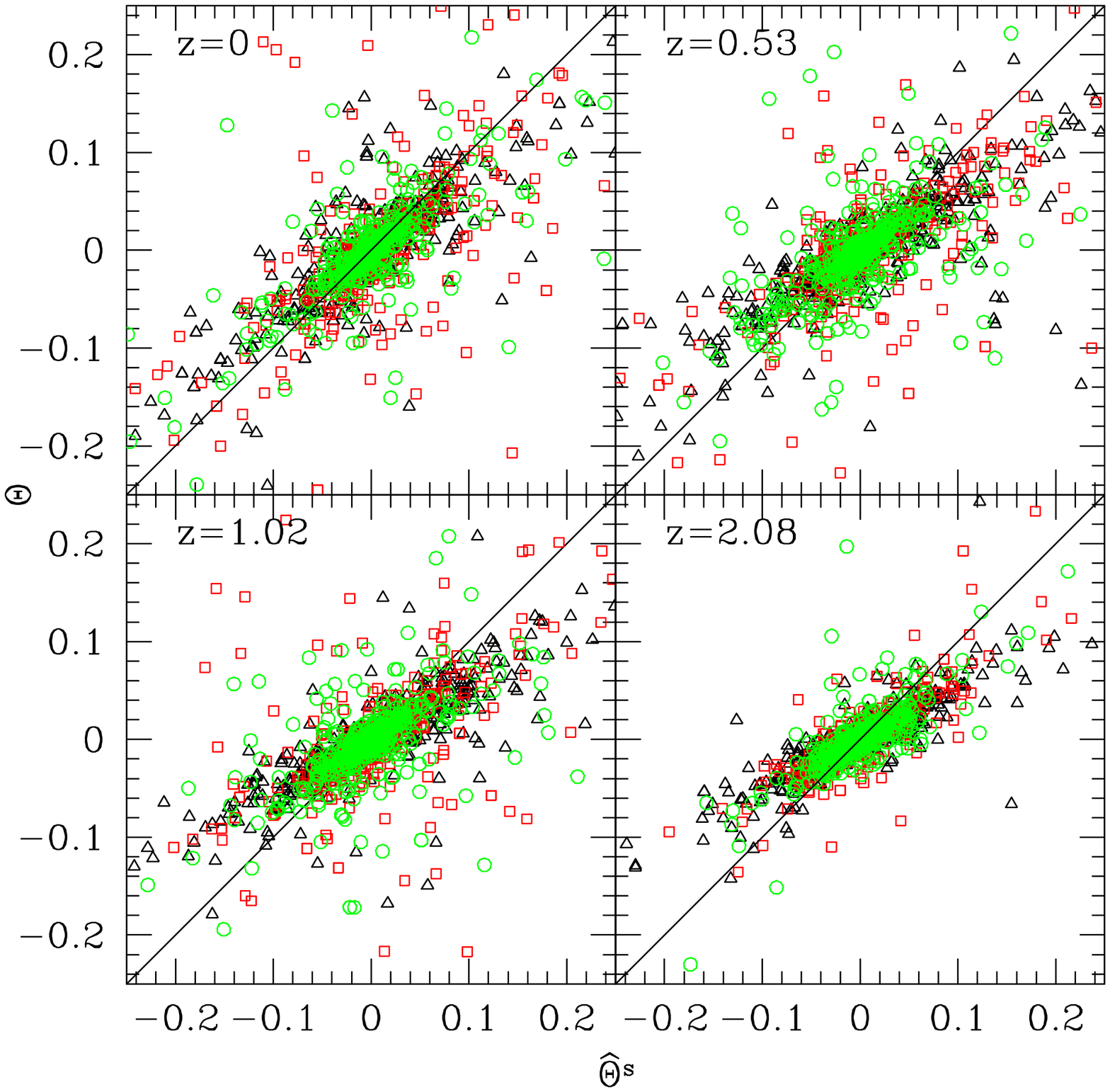}
\caption{The $\Theta$-$\hat{\Theta}^s$ relation in the redshift space. The
same routine as Fig. \ref{fig:rmap64} is used in deriving this relation. At high
redshifts, the recovered $\hat{\Theta}^s$ overweighs the underlying kSZ signal
$\Theta$, because the Kaiser effect enhances the density by a factor
$\sim 1+f/3$, and the higher redshift the higher the slope is, since
$f\simeq \Omega_m^{0.6}(z)$ is larger at early epochs. On the other hand, the
scatters in $\Theta$-$\hat{\Theta}^s$ is much larger than that in the real space
due to the larger stochasticities of density-velocity relation and the real
space-redshift space mapping.}
\label{fig:smap64}
\efi

In reality, what we get from a galaxy survey is the galaxy number density in 
the redshift space. As expected in \S \ref{sec:formalism}, the redshift space
distortion will induce new sources of uncertainty and further degrades the
reconstruction. We are able to quantify its impact through our simulations.
Due to its own peculiar motion, the apparent position of each dark
matter particle along the line of sight can be written as
\begin{equation}
    x^s=x+\frac{{\bf v}\cdot \hat{\bf n}}{H}\ ,
\end{equation}
where $\hat{\bf n}$ denotes the unit vector along the line of sight, and ${\bf
v}$ is the comoving peculiar velocity. ${\bf x}^s$ is position in the redshift
space, and from here on the superscript ``$^s$'' indicates the corresponding
quantity in the redshift space. As a result of this displacement,
what we observed is a distorted density field $\delta^s$. With $\delta^s$ as
the starting point, we can reconstruct the momentum fields $\hat\Theta^s$
through Eq. \ref{eqn:v}, following the same procedure as in the real space.

\subsubsection{$\Theta$-$\hat{\Theta}^s$ relation in the redshift space}
The $\Theta$-$\hat{\Theta}^s$ relation is shown in Fig. \ref{fig:smap64},
which  is significantly different from the one in real space (Fig.
\ref{fig:rmap64}).  We can see that in two aspects. (1) The average slope
of the $\Theta$-$\hat{\Theta}$ relation changes, from $\hat{\Theta}\simeq
\Theta$ to $\hat{\Theta}\simeq a\Theta$ with $a>1$. This is caused by the
linear redshift distortion (the Kaiser effect), which induces \be
\delta^s({\bf k})=\delta ({\bf k}) (1+\beta\mu_k^2),
\label{eqn:lin_z_dis}
\ee
where $\mu_k=\hat{\bf n}\cdot \hat{\bf k}$ is the cosine of the angle between 
the line of sight $\hat{\bf n}$ and the wavevector ${\bf k}$. $\beta=f/b_g$. The
Kaiser effect enhances the galaxy overdensity by a factor $\simeq 1+\beta/3>1$
and thus causes the slope $a>1$. We also notices that the slope $a$ decrease
from high redshifts to low redshifts. There are two causes.  First,
$\beta=f\simeq \Omega^{0.6}_m(a)$ \citep{Peebles1980} (in our case where
$b_g=1$) decreases with decreasing redshift. Second, the finger of God
effect caused by small scale random motion suppresses the redshift space
galaxy density. This effect becomes stronger at lower redshift. (2) Scatters
in the $\Theta$-$\hat{\Theta}^s$ relation are significantly larger than that in
the real space (Fig. \ref{fig:rmap64}). Stochasticities in the density-velocity
relation and real space-redshift space mapping are largely responsible for these
larger scatters. 

\bfi{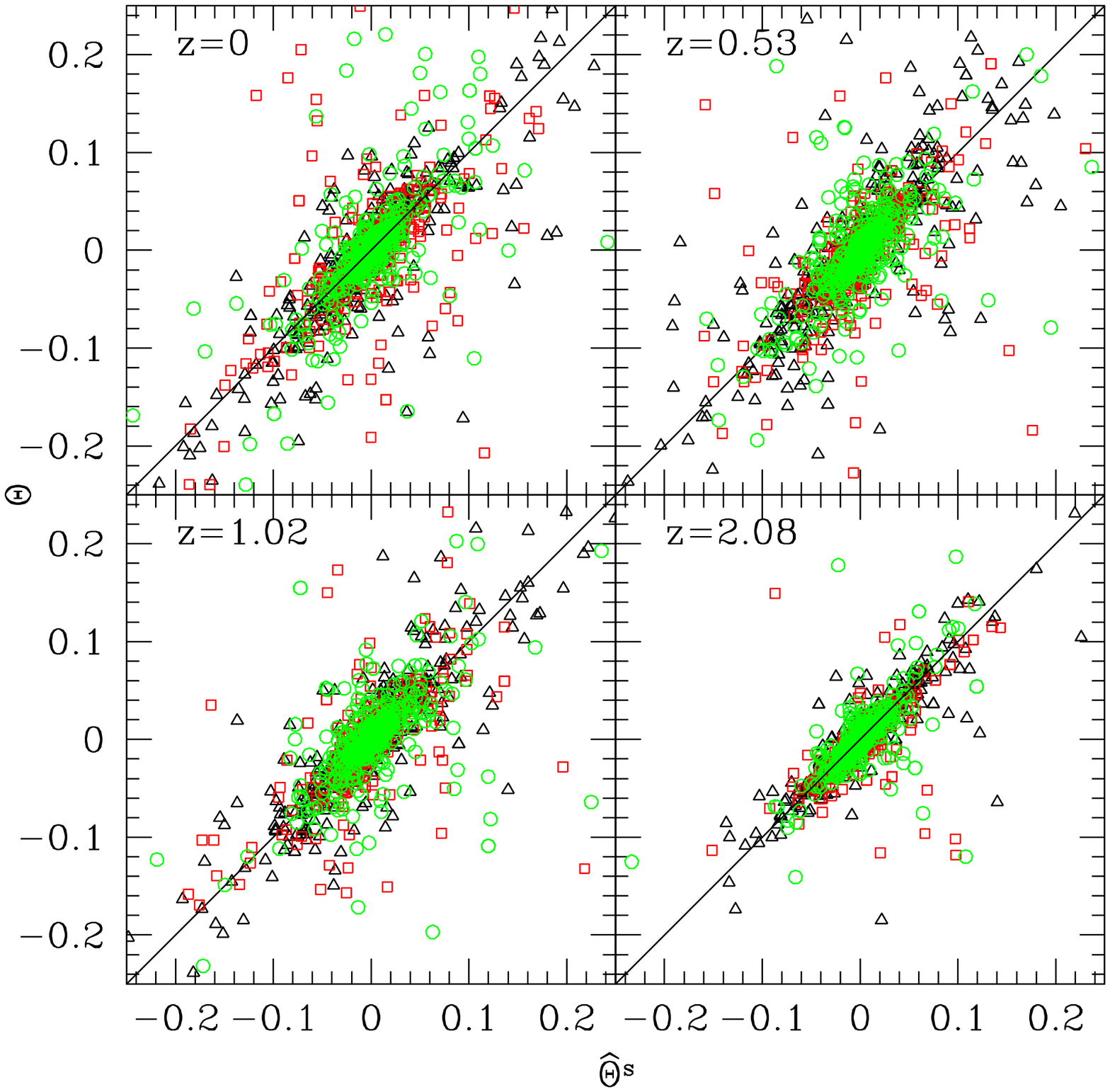}
\caption{The $\Theta$-$\hat{\Theta}^s$ relation in the redshift space after
correcting for the Kaiser effect. It's interesting to find that the slopes
approach unity at all investigated redshifts, while the scatters keep nearly
unchanged compared to Fig. \ref{fig:smap64}.}
\label{fig:s2map64}
\efi

It's interesting to show the $\Theta$-$\hat{\Theta}^s$ relation if we
correct for the Kaiser effect. The figure is shown in Fig. \ref{fig:s2map64}.
We find that the slopes approach unity at all investigated redshifts, while the
scatters keep nearly unchanged compared to Fig. \ref{fig:smap64}. This is,
however, not unexpected, since the deterministic Kaiser formula only changes
the amplitude and is reversible given cosmology. Thus finger of God should be
mainly responsible for the scatters of the relation.

\subsubsection{$r^s(k_\perp,z)$ in the redshift space}
\label{subsec:scorr}
\bfi{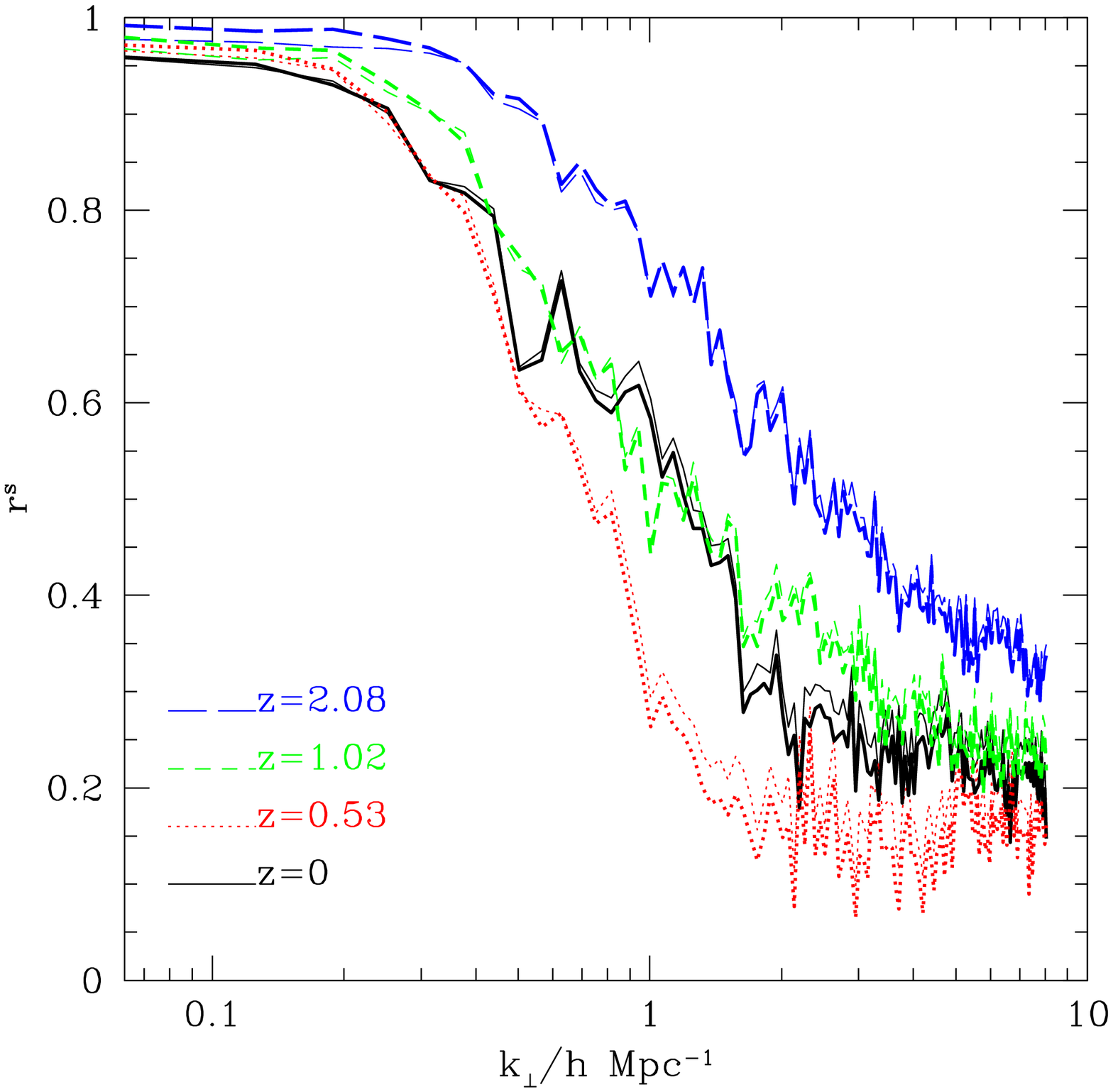}
\caption{The cross correlation coefficients $r(k_\perp,z)^s$ in the redshift
space. Thick lines show $r^s$ with Kaiser effect corrected for.
Compared to Fig. \ref{fig:rcorr}, $r(k_\perp,z)^s$ is suppressed throughout all
scales, especially at $k_\perp\gtrsim1$\hMpc. At $z\lesssim1$, the
reconstruction works poorly, with $r\simeq 0.3$ on scales $k_\perp\gtrsim
3$\hMpc. However, there's a considerable correlation strength $r^s\gtrsim 0.5$
for $k_\perp\lesssim 1$\hMpc at z=0. At higher redshift where nonlinearity of
velocity is not significant, the correlation strength is still tight enough,
e.g. with $r\simeq0.7$ for $k_\perp\sim 1$\hMpc at z=2.08.}
\label{fig:scorr}
\efi

The cross correlation coefficient $r$ in redshift space is shown in Fig.
\ref{fig:scorr}. Compared to Fig. \ref{fig:rcorr}, the correlation coefficient
is suppressed throughout all scales, especially at  $k_\perp\ga 1$\hMpc.  At
$k_\perp \gtrsim 3$\hMpc and $z\la 1$, the kinetic SZ reconstruction and thus
the kinetic SZ tomography works poorly since the reconstructed $\hat{\Theta}$
barely resembles the true signal $\Theta$ ($r\la 0.3$). These results show
unambiguously that redshift distortion is a significant source of error in the
kinetic SZ tomography. It's also worthwhile to see the changes if we
correct for the Kaiser effect. They're also shown in Fig. \ref{fig:scorr} as
thick lines. As indicated by Fig. \ref{fig:s2map64}, correcting for the Kaiser
effect does not influence $r^s$ too much on scales of interest.

As explained in \S \ref{subsec:stochasticity}, the degradation is caused by
the stochasticity in the $\delta$-$\delta^s$ relation. They can be  induced by
the nonlinear mapping between the  real space density and the redshift space
density (e.g.  \citealt{Scoccimarro2004}) and the stochasticity between the
density and velocity field (e.g. \citealt{White2009}). This stochasticity then
induces the stochasticity in the reconstructed velocity $\hat{\bf v}$, with
respect to true velocity (Eq. \ref{eqn:v}). In the momentum reconstruction we
need to multiply the reconstructed velocity by an extra factor $1+\delta$ (Eq.
\ref{eqn:hatT}) to obtain the reconstructed momentum. So the stochasticity in
the momentum reconstruction and thus the kinetic SZ reconstruction has an
extra source, from the term $1+\delta$.  All these complexities worsen the
reconstruction, increase scatters in $\Theta$-$\hat{\Theta}$ and decrease $r$.  

Despite the above degradations, the reconstructed $\hat{\Theta}$ still shows
reasonably tight correlation with the true signal $\Theta$ at $k_\perp\la
1$\hMpc.  We will show in the next section that this correlation strength
allows for robust kinetic SZ tomography, combining the Planck CMB experiment
and the BigBOSS galaxy spectroscopic redshift survey, or other surveys with
comparable power. 

An interesting behavior to notice is that $r^s$ at $z=0.53$ is worse than at
$z=0$. This result is consistent with a larger scatters in the
$\Theta$-$\hat{\Theta}$ plot at $z=0.53$ than at $z=0$, implying that
stochasticities induced by the Finger of God effect becomes the largest at that
epoch. At redshift higher than $z\sim 1$, the reconstruction suffers less from
the nonlinearities and is thus stronger, e.g. with $r>0.7$ at $k_{\perp}<
1h/$Mpc at z=2.08. Actually, we find in simulations that the comoving
velocity dispersion peaks at $z\sim0.6$. This would induce the largest
anisotropies and hence make $r^s$ at $z\sim 0.6$ the worst.

\subsection{Uncertainties in $r$}
In the above section, we quantify the impact of redshift distortion on $r$ and
thus show that redshift distortion degrades  the kinetic SZ reconstruction
significantly. There are other factors affecting the reconstruction. Here, we
briefly discuss the influence of filters adopted to smooth the density field (\S
\ref{subsec:smoothing}) and gastrophysics (\S \ref{subsec:gastrophysics}). 

\bfi{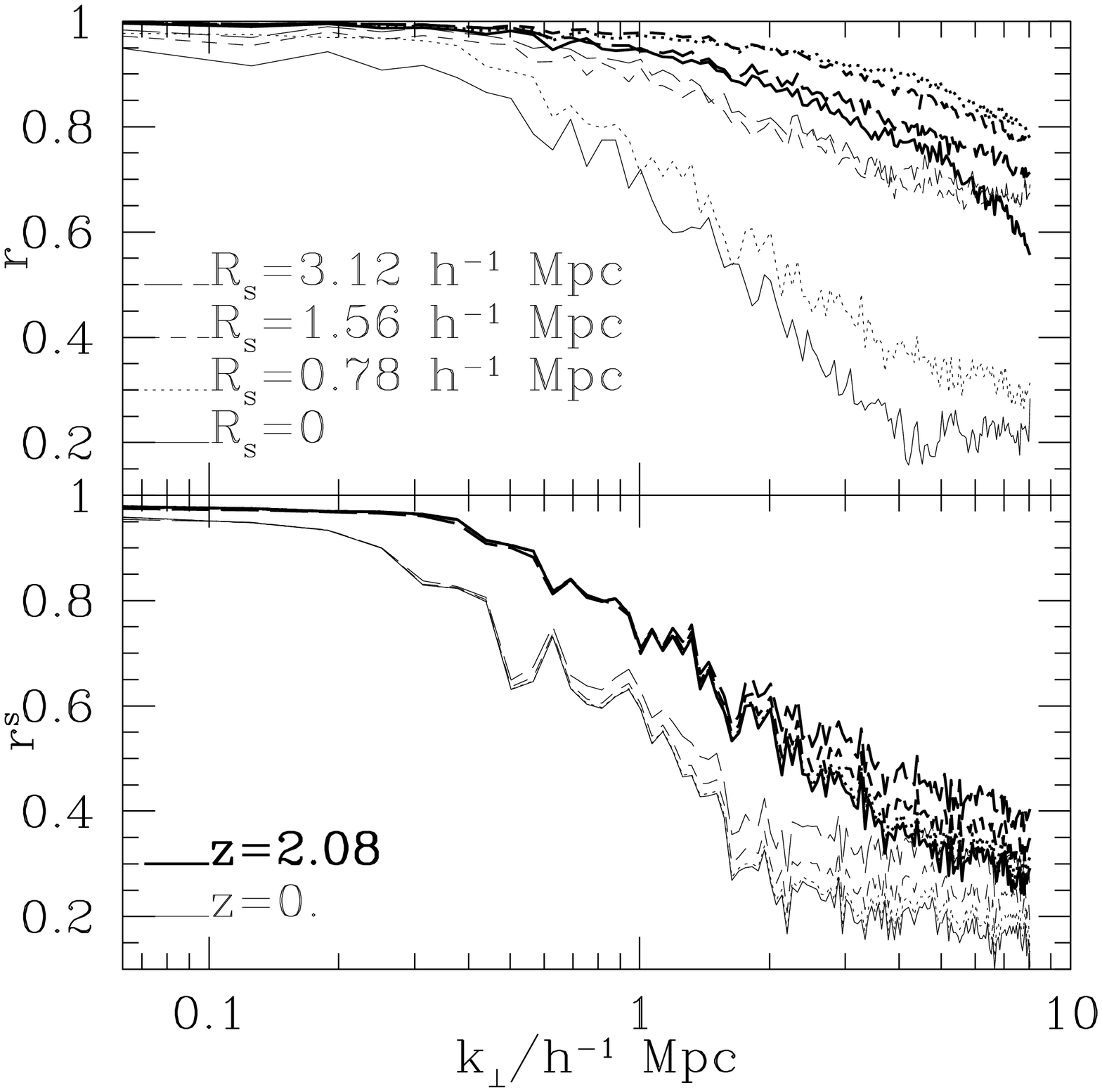}
\caption{The deviations between different smoothing scales, say, 0,
0.78\Mpch, 1.56\Mpch and 3.12\Mpch, in both real space (top panel) and
redshift space (bottom panel). In the real space, smoothing length with
0.78\Mpch works best at z=2.08 while 3.12\Mpch is the most suitable at z=0.
However, in the redshift space, the enhancement due to smoothing is less
effective, and a larger smoothing length is better. Probably an anisotropic
smoothing other than a Gaussian smoothing should work better. As a median
smoothing length, $1.56$\Mpch is effective at both low and high redshift.
}
\label{fig:rdiv}
\efi

\bfi{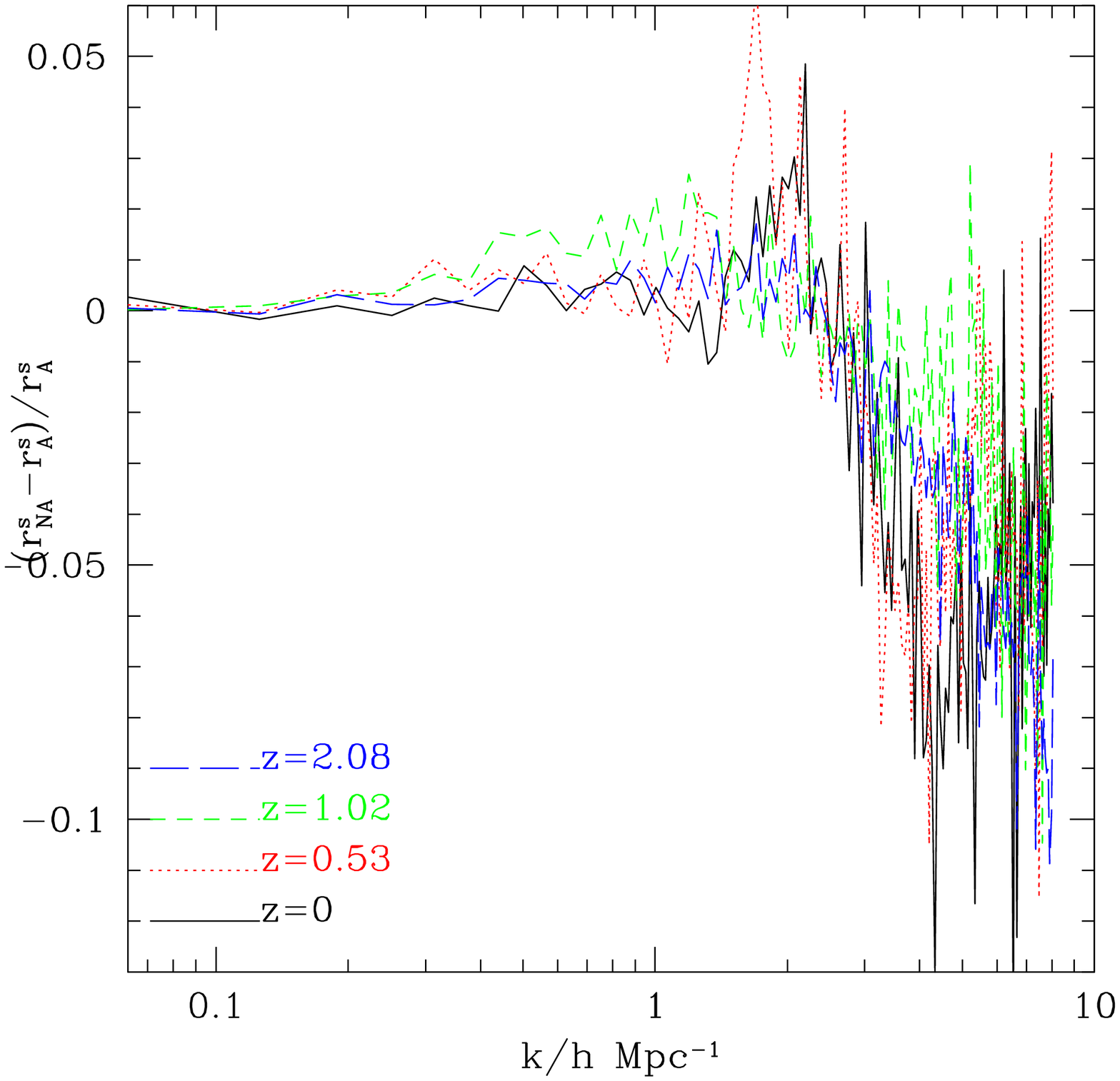}
\caption{The difference of the correlation coefficient between adiabatic and
non-adiabatic simulations. The overall $r^s$ is insensitive to gastrophysics,
and there're only $\sim$2-3\% changes of $r^s$. It will accordingly influence
the estimator of the signal-to-noise of the cross power spectrum by several
percent level, however, for kSZ tomography, this is sufficient for the current
surveys.}
\label{fig:simulation}
\efi

\subsubsection{The influence of density smoothing on $r$}
\label{subsec:smoothing}
The results shown above all adopt  a Gaussian filter with smoothing length
$R_s=1.56$\Mpch to smooth the density field before velocity reconstruction.
The reconstruction robustness no wonder depends on the way to smooth the
density field. We do not aim to perform a comprehensive investigation on this
issue. Rather, we will restrict to the Gaussian filter and investigate the
dependence of $r$ on $R_s$. We arbitrarily compare between the cases of
$R_s=0,0.78, 1.56,3.12$ \Mpch.  For clarity, we only show the comparisons at
$z=0$ and $z=2.08$ in Fig. \ref{fig:rdiv} . 

The basic (and obvious) conclusion is that, smoothing is necessary to suppress
small scale nonlinearities and improve the reconstruction. Fig. \ref{fig:rdiv}
(upper panel) shows general improvement in $r$ when smoothing is taken,
comparing to the case of no smoothing ($R_s=0$). For example, smoothing the
density field with $R_s=1.56$\Mpch can boost $r$ by $\sim 30\%$ at $k=1$\hMpc
and a factor of $2$ or more at smaller scales, in real space. 

Gaussian smoothing is less effective in redshift space: the improvement in
$r_s$ is often less than  $10\%$ on scales of interest (bottom panel, Fig.
\ref{fig:rdiv}).  The redshift space overdensity is anisotropic, so a
spherical Gaussian smoothing will not work well. An anisotropic filter may
work better in redshift space. This is certainly an interesting technical
issue for further investigation. 

A right smoothing should balance between suppressing small scale nonlinearities
and preserving large scale signal. If $R_s$ is too large, it may wipe off too
much large scale clustering responsible for peculiar velocity and thus degrade
the reconstruction (decrease $r$). This may be the reason that $R_s=0.78$\Mpch
works better than larger $R_s$, at $z=2.08$ in the real space. However,
overall $R_s=1.56$ \Mpch works reasonably well at all redshifts investigated.  

The simulation data we deal with is like an ideal survey, with negligible shot
noise, uniform selection function and regular survey boundary.  Smoothing for
real data is of course much more complicated. For example, a big issue in real
survey is shot noise due to low galaxy number density, especially in
spectroscopic redshift surveys. For this issue, one can refer to
\citet{Ho2009} for discussion on the application of the Wiener filter.  

\subsubsection{$r$ and gastrophysics}
\label{subsec:gastrophysics}
$r$ also depends on gastrophysics. All the results shown above are based on
our non-adiabatic simulation, with radiative cooling, star formation and
supernova feedback. Although we are not able to robustly quantify its
detailed dependence on these gastrophysical processes, we can obtain a rough
estimation by comparing the above results to our adiabatic simulation with
identical initial conditions. The relative differences of $r^s$ between the two
simulations are shown in Fig. \ref{fig:simulation}. We find that, overall $r$
and thus the performance of the kinetic SZ tomography is insensitive to the
gastrophysics. $r$ at $k_{\perp}<1$\hMpc only varies by less than
$2$-$3\%$. The influence of gastrophysics is larger at smaller scales, but is
still less than $5\%$ up to $k_{\perp}=3$\hMpc. 

This insensitivity to gastrophysics has two implications on our kinetic SZ
tomography. First, it is unlikely that some realistic gastrophysical process
not included in our non-adiabatic simulation can dramatically suppress $r$ and
thus invalidates the tomography.  Second, this significantly simplifies the
theoretical interpretation of the tomography results.  Based on the same
technique in \citep{Shao2009} , we can combine the two {\it measured}
correlations, $\langle \Theta\hat{\Theta}\rangle$ and $\langle \hat{\Theta}
\hat{\Theta}\rangle$, to obtain $\langle \Theta\Theta\rangle$ arising from the
same redshift bin. The only unknown quantity in this approach is $r$, which we
shall calibrate against simulations.  However, if $r$ is very sensitive to
gastrophysics, the calibration on $r$ would be very difficult due to large
uncertainties in  our theoretical and numerical understanding of these
gastrophysics. The insensitivity of $r$ to gastrophysics shown in Fig.
\ref{fig:simulation} implies that, despite imperfect theoretical and numerical
understanding of  these gastrophysics, $r$ can still be accurate to a few
percent level at relevant scales. This precision suffices for the kinetic SZ
tomography. 

\bfi{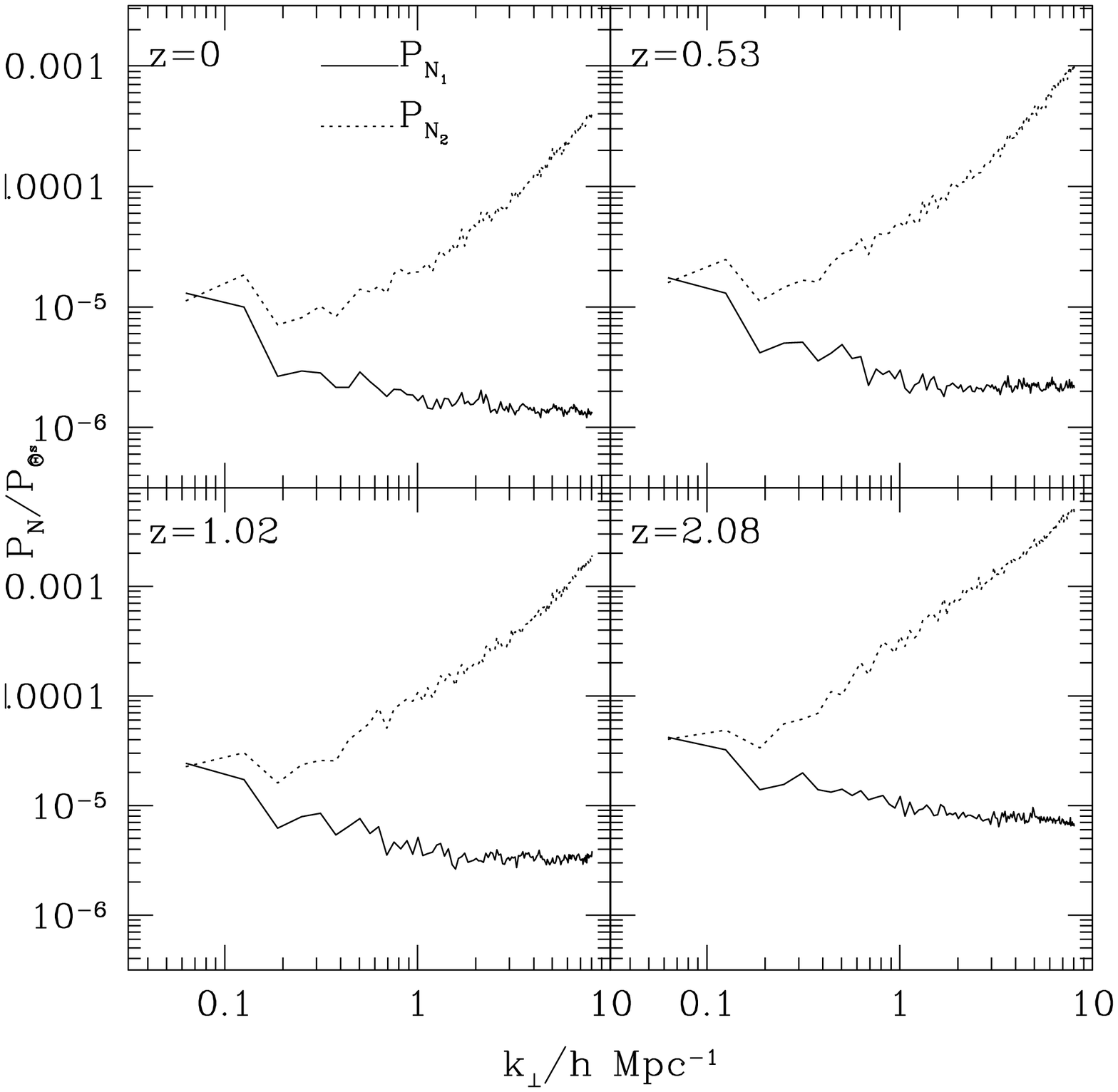}
\caption{The ratio $P_{N_{1,2}}/P_{\hat{\Theta}^s}$ of the shot noise power
spectrum and the reconstructed momentum power spectrum. Both the
$\Delta^2_{N_1}$ and $\Delta^2_{N_2}$ are much smaller than the reconstructed
momentum powers, with $N_2$ terms dominating $N_1$ towards smaller scales.
Given a small number density in current galaxy redshift survey, the total shot
noise would be comparable to the signal. Nevertheless, on scales of interest
for kSZ tomography, say $k_\perp \lesssim 1$\hMpc, galaxy surveys would
provide a relatively less contaminated reconstruction of the momentum.} 
\label{fig:pown}
\efi

\section{Error forecast}
\label{sec:err_fore}
The $r$ measured above quantifies the performance of the kinetic SZ tomography
for a virtually ideal galaxy survey, for which the simulation particle number density
is high and the shot noise is negligible. However, in real surveys, the galaxy number
density is a factor of $10^4$-$10^5$ smaller, resulting in much larger and thus
non-negligible shot noise. The shot noise affects both the velocity and momentum
reconstruction. We use our simulation to quantify these effects and then apply
these results to forecast the performance of the kinetic SZ tomography. Our
target CMB experiment is Planck and the target galaxy spectroscopic survey is
BigBOSS \citep{BigBOSS2009}. The kinetic SZ tomography based on other SZ surveys like SPT and
galaxy surveys like ADEPT, Euclid and SKA is expected to work better.

\subsection{Estimating the reconstruction noise from galaxy distribution shot noise}
\label{subsec:shotnoise}
The discreteness of dark matter particles induces spurious fluctuations (shot
noise)  in the density field, which we denote as $\delta_N$, and thereby  in the
reconstructed velocity, which we denote as ${\bf v}_N$.  The correlation of
the contaminated momentum $\hat{\Theta}_c$ is (for simplicity, we omit the
projection along the line of sight) 
\ba
\langle\hat{\Theta}^s_c(\hat{n})W\hat{\Theta}^s_c(\hat{n}
^\prime)\rangle&\propto& \langle (1+\delta^s+\delta_N)(\hat{\bf v}^s+\hat{\bf
v}_N)\cdot\hat{n} \nonumber  \\
 && \times (1+\delta^{s\prime}+\delta_N^\prime)(\hat{\bf
v}^{s\prime} +{\bf v}_N^\prime)\cdot\hat{n}^\prime  \rangle \nonumber\\
  &=&\langle(1+\delta^s)\hat{\bf v}^s\cdot\hat{n} (1+\delta^{s\prime})\hat{\bf
v}^{s\prime}\cdot\hat{n}^\prime \rangle \nonumber \\
 &&+   \langle (1+\delta^s) {\bf v}_N\cdot\hat{n}  (1+\delta^{s\prime}) {\bf
      v}_N^\prime\cdot\hat{n}^\prime \rangle \nonumber\\
    &&+\langle \delta_N \hat{\bf v}^s\cdot\hat{n}
\delta_N^\prime \hat{\bf v}^{s\prime}\cdot\hat{n}^\prime \rangle\ . \nonumber
\ea
The last two terms in the above equation are the leading non-vanishing
noise terms. We denote the first noise term as $N_1\propto (1+\delta^s){\bf
v_N}$ while the second as $N_2\propto \delta_N \hat{\bf v}^s$

By randomizing the particle positions in the simulation, we can directly
measure $\delta_N$ and derive therein ${\bf v}_N\propto \delta_N(k)\hat{k}/k$.
Following the same procedure as in reconstructing the momentum map, we produce
two corresponding maps of these two shot noises. In order to estimate the
contribution of shot noise, we compute the power spectrum of the two noise
terms, and show the ratios $P_{N_{1,2}} /P_{\hat{\Theta}^s}$ in Fig.
\ref{fig:pown}. We can see that both the $P_{N_1}$ and $P_{N_2}$ noise terms
are much smaller than the reconstructed momentum powers. Although $N_1$ and
$N_2$ terms are comparable on the very large scales, they deviate
significantly towards smaller scales, with $N_2$ terms predominantly
overweighing $N_1$ terms. On the small scales, $N_2$ terms would turn out to
be several thousandth of the reconstructed momentum $\hat{\Theta}^s$, and they
can be comparable to or even dominate the signal, given a much smaller number
density in current galaxy redshift surveys.  Nevertheless, on scales of
interest for kSZ tomography, say $k_\perp \lesssim 1$\hMpc, galaxy surveys
would provide a relatively less contaminated reconstruction of the momentum.  

\subsection{Error forecast in $C_l$}
Based on the above results, we are able to estimate the S/N for real
surveys. Since in this section we  consider the error forecast in real survey,
i.e. in the redshift space, we omit the superscript ``$^s$'' for simplicity.
For a given redshift bin $\in[z-\Delta z/2,z+\Delta z/2]$, we are able to
measure the angular cross power spectrum $C_{\ell}$ between the
reconstructed $\hat{\Theta}$ and the CMB measurement where the kSZ effect is
embedded.
The statistical error in this measurement can be estimated by 
\ba
\label{eqn:cl}
    \frac{\Delta C_{\ell}}
{C_{\ell}}&\simeq&
\sqrt{ \frac{1+\frac{C_\ell^{\rm CMB}+C_\ell^{\rm kSZ}+C_\ell^{\rm
CMB,N}}{{r}^2 C_\ell^{\rm kSZ,\Delta z} }(1+\frac{C^{N}_{\hat{\Theta}}}
{C_{\hat{\Theta}}})} {2\ell\Delta \ell f_{\rm sky}}} \\
    &\simeq & \frac{1}{2r\ell\Delta \ell f_{\rm sky}} \sqrt{ \frac{C_\ell^{\rm
CMB}+C_\ell^{\rm kSZ}+C_\ell^{\rm CMB,N} } {C_\ell^{\rm kSZ,\Delta z}}
\left(1+\frac{C^{N}_{\hat{\Theta}}}{C_{\hat{\Theta}}}\right)} \ . \no
\ea
Here, $C_{\ell}^{\rm CMB}$ is the angular power spectrum of primary CMB and
$C_{\ell}^{\rm CMB,N}$ is the power spectrum of the measurement noise. The
thermal SZ effect is also a source of noise. Since we work with the frequency
band around 217 Ghz, where the thermal SZ effect virtually vanishes, we will
neglect the thermal SZ effect. There are other possible sources of error, such
as the dusty star forming galaxies and radio sources. They are not likely
dominant over the primary CMB at $\ell\la 2000$ of our interest. Thus  we will
not include them in the error analysis. $C_{\ell}^{\rm kSZ}$ is the angular
power spectrum of the kinetic SZ effect and $C_{\ell}^{\rm kSZ, \Delta z}$ is
the contribution from the given redshift bin. $C^N_{\hat{\Theta}}/
C_{\hat{\Theta}}$ is the ratio of the angular power spectrum of the
reconstruction shot noise and of the reconstructed momentum
$\hat{\Theta}$. $r$ is again the cross correlation coefficient between the
reconstructed kSZ map and the true kSZ signal.  However, now the projection
length is in general much larger than the simulation box size $100h^{-1}$Mpc.
For interesting angular scales of $\ell \sim 10^3$, two redshift bins
separated by more than $100h^{-1}$Mpc can be approximated as uncorrelated. So
we can average over $r$ measured in the last section over the relevant
redshift range to obtain an estimation of the $r$ used in this section. 

The first factor $1$ in the r.h.s of Eq. \ref{eqn:cl} comes from the cosmic
variance in the cross-correlation signal and assumes the kinetic SZ effect and
the reconstructed $\hat{\Theta}$ are Gaussian. The actual non-Gaussianties
will increase the cosmic variance. However, this cosmic variance term is
sub-dominant to the other term, since $(C_\ell^{\rm CMB}+C_\ell^{\rm
kSZ}+C_\ell^{\rm CMB,N})/ C_\ell^{\rm kSZ,\Delta z}\gg 1$ and $r\leq 1$. So
the Gaussian assumption is reasonable. And the second relation holds since
usually $(C_\ell^{\rm CMB}+C_\ell^{\rm kSZ}+C_\ell^{\rm CMB,N})/ C_\ell^{\rm
kSZ,\Delta z}\gg 1$. 

From this equation, we can figure out the improvement in S/N of the kinetic SZ
measurement. To better see this point, we will discuss under the
limit that the galaxy number density is sufficiently high such that
$C^{N}_{\hat{\Theta}}/C_{\hat{\Theta}}\ll 1$ at sufficiently large angular
scales. We then have
\be
\frac{S}{N}\sim \sqrt{\frac{r\times 2\ell\Delta \ell f_{\rm sky} C_\ell^{\rm
kSZ,\Delta z}} {C_\ell^{\rm CMB} + C_\ell^{\rm kSZ}+C_\ell^{\rm CMB,N}}}\ .
\ee
We can compare it to the S/N of the kSZ auto correlation power spectrum
measurement
\be
\frac{S}{N}\sim \sqrt{\ell\Delta
    \ell f_{\rm sky}}\frac{C_\ell^{\rm kSZ}}{C_\ell^{\rm CMB}+C_\ell^{\rm
    kSZ}+C_\ell^{\rm CMB,N}}
\ee
As $C_\ell^{\rm CMB}\gg C_\ell^{\rm kSZ}$ at $\ell\la 2000$, the
improvement in S/N by the kSZ tomography is of the order $r\sqrt{C_\ell^{\rm
CMB}/C_\ell^{\rm kSZ}}\sqrt{C_\ell^{\rm kSZ,\Delta z}/C_\ell^{\rm kSZ}}$.
Since $r$ is close to unity at relevant scale (e.g. $\ell\sim 10^3$ and $z\sim
1$), for a sufficiently deep galaxy survey (e.g. to $z=2$ or beyond), the
overall improvement is of the order $\sqrt{C_\ell^{\rm CMB}/C_\ell^{\rm
kSZ}}\gg 1$. In reality, since the galaxy number density is finite, shot noise
dominates at small scales, as can be inferred from Fig. \ref{fig:pown}.

\bfi{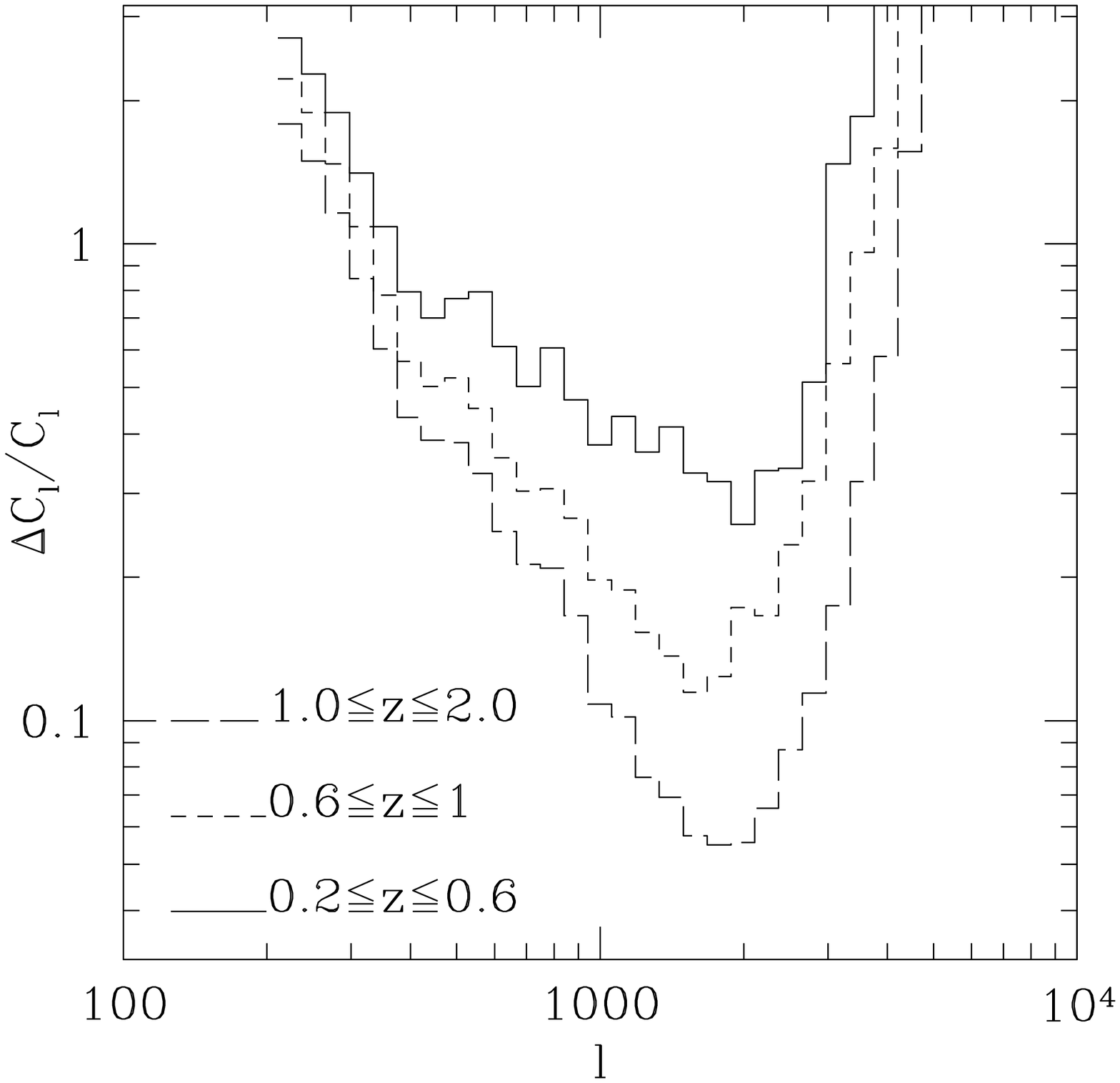}
\caption{The error of the measurements of cross power spectrum between at
redshift bins z=$0.2-0.6$, $0.6-1.0$ and $1.0-2.0$, using BigBOSS plus PLANCK.
There are significant signals around $\ell\sim 2000$, with errors 26\%,
11\% and 6\% respectively for the three redshift bins. With a larger redshift
bin and better $r^s$ over the relevant multipole ranges, the total S/N in
$z=1.0-2.0$ is 46, the more considerable than the other two, 10 at z=0.2-0.6 and
21 at z=0.6-1.0.}
\label{fig:snr}
\efi

Now we numerically calculate the S/N for the combination of PLANCK plus
BigBOSS-N. $C_{\ell}^{\rm 
CMB}$ is calculated from CAMB\footnote{http://camb.info/}. 
The instrument noise power spectrum of PLANCK is 
\be
C_\ell^{\rm CMB,N}=(\sigma_{p,T} \theta_{\rm FWHM})^2 W_\ell^{-2}\ ,
\label{eqn:cmb_noise}
\ee
where the window function $W_\ell=\exp(-\ell(\ell+1)/2\ell^2_{\rm beam})$ for
a Gaussian beam with $\ell_{\rm beam}=\sqrt{8\ln{2}}/ \theta_{\rm
FWHM}$.\footnote{http://www.rssd.esa.int/index.php?project=planck} For PLANCK,
$\theta_{\rm FWHM}=5.0$ arcminutes. Within which the average 1$\sigma$
sensitivity per pixel $\sigma_{p,T} /T_{\rm CMB}=4.8\times 10^{-6}$ is
expected after 2 full sky surveys. PLANCK will survey the sky at several
frequency bands. The  217 GHz band where thermal SZ signal vanishes is the one
that we use to cross correlate with galaxies.\footnote{The $217$ GHz band is
the only band promising to detect the kinetic SZ effect in auto correlations.
However, the kSZ tomography is able to detect the kSZ effect in other
frequency bands and the resulting S/N can be comparable at= $\ell<2000$  where
the thermal SZ effect is sub-dominant to the primary CMB. This is an issue for
further investigation. } $C_{\ell}^{\rm kSZ}$ and $C_{\ell}^{\rm kSZ,\Delta
z}$ are calculated by the model of \citet{Zhang2004}. 

For the target galaxy spectroscopic redshift survey, we choose the BigBOSS(-N)
project \citep{BigBOSS2009}. It plans to measure the spectroscopic redshifts
of 30 million LRGs and emission line galaxies at $0.2\le z \le 2$ over 14000
deg$^2$ sky coverage. Since PLANCK  will survey  the full sky, the overlapping
fractional sky coverage is $f_{\rm sky}=14000/(4\pi/(\pi/180)^2)$. Given the
much lower galaxy number density (comparing to simulations), shot noise is a
significant source of error in the kinetic SZ tomography. The resulting
reconstruction noise can be roughly estimated through a
simple scaling
 \be
\frac{C^N_{\hat{\Theta}}(\ell)}{C_{\hat{\Theta}}(\ell)}
\sim \frac{P_{N_1}(k_\perp)+P_{N_2}(k_\perp)} {P_{\hat{\Theta}}(k_\perp)}
\frac{\Sigma_{\rm sim}}{\Sigma_{\rm obs}}\ ,
\ee
where, $P$s are the ones at some intermediate redshift of the redshift bin and
$k_{\perp}=\ell/\chi$.  $\Sigma_{\rm sim}$ and $\Sigma_{\rm obs}$ are the surface
densities in the simulation and in the real survey respectively. This
estimation is by no means exact. But it is reasonably good to demonstrate the
power of PLANCK plus BigBOSS to detect kSZ. 

We divide the BigBOSS survey into three redshift bins,
$0.2\le z \le 0.6$, $0.6\le z \le 1.0$ and $1.0\le z\le2.0$. This choice is
somewhat arbitrary. In the error estimation, we simply adopt $r$ and
$C_{\hat{\Theta}}^N/C_{\hat{\Theta}}$ estimated at some intermediate redshift
of each redshift bin. In the error estimation using Eq. \ref{eqn:cl}, the
multipoles $\ell$ are logarithmically equal spaced, say, $\Delta\ell/\ell=12.2\%$.
We present the error distribution of the cross power spectrum in Fig.
\ref{fig:snr}.  We find that the cross correlation can be measured robustly at
$\ell\sim 1000$-$3000$. For example,  the relative errors at $\ell=2000$ are
about 26\%, 11\% and 6\% at the three redshift bins correspondingly. Good
performance at these scales is expected for three reasons. (1) $\hat{\Theta}$
and the underlying $\Theta$ are tightly correlated at corresponding scale
$k_{\perp}=\ell/\chi$. (2)  Shot noise is low for a galaxy survey as big as
BigBOSS, comparing to the noise level in the kSZ measurement. (3) The primary
CMB drops  while the measurement shot noise is still bearable  at $\ell=2000$. 

The accuracy of the cross power spectrum measurement increases with redshift.
This is partly due to stronger $r$ at higher redshift. And this epoch, the
comoving scale $k_{\perp}=\ell/\chi(z)$ moves to more linear regime and the
nonlinearity becomes weaker. Both push $r$ upward and improve the kSZ
tomography. What's more, the binned kSZ signal from z=1.0-2.0 is larger than
the other two on scales of interest. It certainly helps the cross correlation
measurement.

On the other hand, it is unlikely to detecting kSZ at $\ell\la 300$ and
$\ell\ga 3000$ through the kSZ tomography with PLANCK plus BigBOSS-N . Toward
larger scales, the kSZ signal decrease, while the primary CMB overweighs by
3 or 4 orders. On smaller scales, the finite angular resolution of the PLANCK
survey causes the CMB measurement noise to increase exponentially. 
At the same time, the shot noise in BigBOSS begins to dominate over the galaxy
clustering signal.  Here we want to caution the readers that the S/N at
$\ell\ga 3000$ is overestimated, since we neglected noises from the dusty star
forming galaxies \citep{Hall2010}, which may dominate over the primary CMB. 

We estimate that, the overall S/Ns for the $z=0.2-0.6$, $0.6$-$1.0$ and
$1.0$-$2.0$ redshift bins are  $10$, $21$ and $46$, respectively.  The
combined S/N of the  whole redshift range $z=0.2-2.0$ is  $51$. These results
are consistent with the findings of \citet{Ho2009},  in which they use PLANCK
plus SDSS and ADEPT.  Although these numbers are likely only accurate within a
factor of 2,  they nevertheless robustly confirm the applicability of the proposed kSZ
tomography to realistic surveys. Since the overall S/N$\sim 50$, we are able
to choose finer redshift bins and thus measure the kSZ evolution over $\sim
10$ redshift bins.

\section{Discussions and conclusions}
\label{sec:conclusion} 
In this paper, we proposed the kSZ tomography method, tested it against
simulations and estimated its applicability to realistic surveys. This methods
requires a galaxy spectroscopic redshift survey to reconstruct the large scale
peculiar velocity and then weigh the galaxy density with this velocity to
produce a weighted galaxy momentum map $\hat{\Theta}$. By construction, this
map is tightly correlated with the true kSZ signal and thus allows us to
extract the kSZ signal from the CMB maps through the cross correlation
measurement. Our work confirms the major finding of \citet{Ho2009}, namely
cross correlating the properly  weighted galaxy momentum field with CMB
fluctuations can significantly improve the kinetic SZ measurement.  We
estimate that BigBOSS-N plus PLANCK is able to measure the kSZ effect at  $50
\sigma$ confidence level, qualitatively consistent with results  of
\citet{Ho2009}.  Given this S/N,  our method is able to recover the redshift
distribution of the kSZ effect over $\sim 10$ redshift bins.  

Thanks to the hydrodynamical simulations at hand, we are able to better model
the true kSZ signal and thus reduce one source of error in the analysis. We
are also able to quantify the impact of redshift distortion, by performing the
reconstruction in both real and redshift space. We find that redshift
distortion is a major factor affecting the kSZ tomography. It degrades the
performance, especially at small scales.  However, we show that, its impact
at $k_{\perp}\la 1h/$Mpc is moderate and thus the kSZ  tomography can still
work reasonably well to scales $\ell\simeq k_{\perp}\chi\sim 3000$. 

Comparing between the simulations with star formation, gas cooling and
supernovae feedback turned on or off, we are able to quantify the impact of
gastrophysics.
We showed that the correlation coefficient $r$ between the reconstructed
$\Hat{\Theta}$ and the true kSZ signal $\Theta$ is insensitive to
gastrophysics of star formation, feedback and gas cooling included in our
hydrodynamic simulations. This behavior is of crucial importance for the
theoretical interpretation of the kSZ tomography result.  It enables an
approach similar to the thermal SZ tomography \citep{Shao2009} to circumvent
the problem of the potentially (and likely inevitably) large bias in
$\hat{\Theta}$ with respect to $\Theta$. This is an issue for further
investigation. 

The proposed kSZ tomography has unique advantages over detecting the kinetic
SZ effect through auto-correlations. (1) By construction, the cross
correlation measurement only picks up the kSZ component in CMB. Due to the
lack of characteristic directional dependence,  primary CMB, the thermal
SZ effect, dusty star forming galaxies and any noise sources of scalar nature
do not bias the cross correlation measurement. It is thus a rather clean way
to eliminate the otherwise overwhelming systematical errors in the kSZ
measurement.  (2) It recovers the redshift information of the kSZ effect. The
redshift information is not only useful to better understand the evolution of
missing baryons, but also useful to separate the kSZ effect after reionization
from the one due to patchy reionization. It thus allows for better understanding
of the reionization process. (3) The relatively higher S/N of galaxy surveys
help to beat down the statistical errors in the kSZ measurement, as we have
shown for the combination of PLANCK plus BigBOSS-N. 

We have carried out a concept study and shown the applicability of the kSZ
tomography to real surveys. However, the analysis is simplified, with many
issues left for future study. An incomplete list is as follows. 
\begin{enumerate}
\item The galaxy stochasticity. It definitely degrades the kSZ tomography
  performance. However,  studies shown that the
  stochasticity in the galaxy bias is likely at the level of $10\%$ or smaller
  \citep{BP2009,Baldauf2010}.  So it is unlikely to completely invalidate the
  kSZ tomography, consistent with \citet{Ho2009} in which the
  stochasticity exists in their galaxy mock catalogue. Nonetheless, it is
  important to quantify the impact of galaxy stochastic bias. 
\item The smoothing scheme. The actual galaxy density distribution is irregular, due
  to the intrinsic nonlinearities, galaxy shot noise and survey
  irregularities. So an important step is to smooth the density field and
  stabilize the velocity reconstruction.  We have done simple study on the
  smoothing scheme in \S \ref{subsec:smoothing} and found that this is indeed
  important.  For example, we find that a spherical Gaussian smoothing does
  not work in redshift space as well as in real space. \citet{Ho2009}
  investigated the Weiner filter to deal with galaxy shot noise. A
  comprehensive study is required to develop an optimal smoothing scheme. 
\item The cosmic variance.  Our simulated box size is not large enough to
  quantify the cosmic  variance,  as the correlation length of velocity is
  $\sim 100$ \Mpch.  Simulations with larger box size are required to study this
  issue, along with the galaxy stochasticity and smoothing scheme. 
\item The robustness of modeling $r$, the cross coefficient $r$ between the
  reconstructed $\Hat{\Theta}$ and the true kSZ signal $\Theta$. It plays a
  central role in quantifying the performance of the kSZ tomography and in
  converting the measured cross correlation signal to the kSZ auto correlation
  power spectrum. We have measured the impact of gastrophysics on $r$ and
  found that, gastrophysics of star formation, supernovae feedback and gas
  cooling would only change $r$ by several percent on scales of interest. We
  shall check this result against more comprehensive investigations on
  gastrophysics.  We have measured the dependence of $r$ on the smoothing length
  and found a strong dependence. This implies that $r$ in real surveys which
  have complicated masks and noise distribution could differ from what we have
  presented here.  
\item Improvements on the forecast.  The forecast can be improved by including
other sources of contamination such as dusty star forming galaxies. It can
also be extended to other combinations such as CMB   experiments like ACT and
SPT and many other galaxy spectroscopic redshift surveys. Dusty star forming
galaxies do not have significant effect on the kSZ tomography based on  PLANCK
plus BigBOSS, since the applicable scale is $\ell\la 3000$. At smaller scales
approachable to ACT and SPT, dusty star forming galaxies overwhelm both the
primary CMB and the thermal and kinetic SZ effect. So for these surveys, this
source of error must be taken into account appropriately. 
\item Probing missing baryons with the kSZ tomography. Although it is one of
  major motivations and the most important applications of the kSZ tomography,
  we have not carried out any quantitative analysis on this aspect. Our hydro
  simulations have rich information on the IGM. However, the simulation box is
  not sufficiently large to robustly quantify the contribution of the missing
  baryons to relevant statistics in the kSZ tomography. Furthermore,  more
  comprehensive and robust treatments on gasdynamics and galaxy formation  are
  demanded to reveal the connections between the  kSZ effect and density and
  velocity of galaxies. Again, this requires more hydrodynamical simulations,
  which will be investigated elsewhere.  
\end{enumerate}

\section{Acknowledgement}
We thank S. Ho for his insightful questions and suggestions to improve the
paper. This work is supported in part by the National Science Foundation of
China (grant No.  10533030, 10673022, 10821302, 10873027\&10878001), the
Knowledge Innovation Program of CAS (grant No. KJCX2-YW-T05 \& KJCX3-SYW-N2),
the 973 program grant No. 2007CB815401 \& 2007CB815402 and the CAS/SAFEA
International Partnership Program for Creative Research Teams. The simulations
were done at Shanghai Supercomputer Center by the supports of Chinese National
863 project (grant No.06AA01A125).

\appendix

\section{The reconstruction bias $b_{\hat{\Theta}}$}	
\label{sec:app1}
\bfi{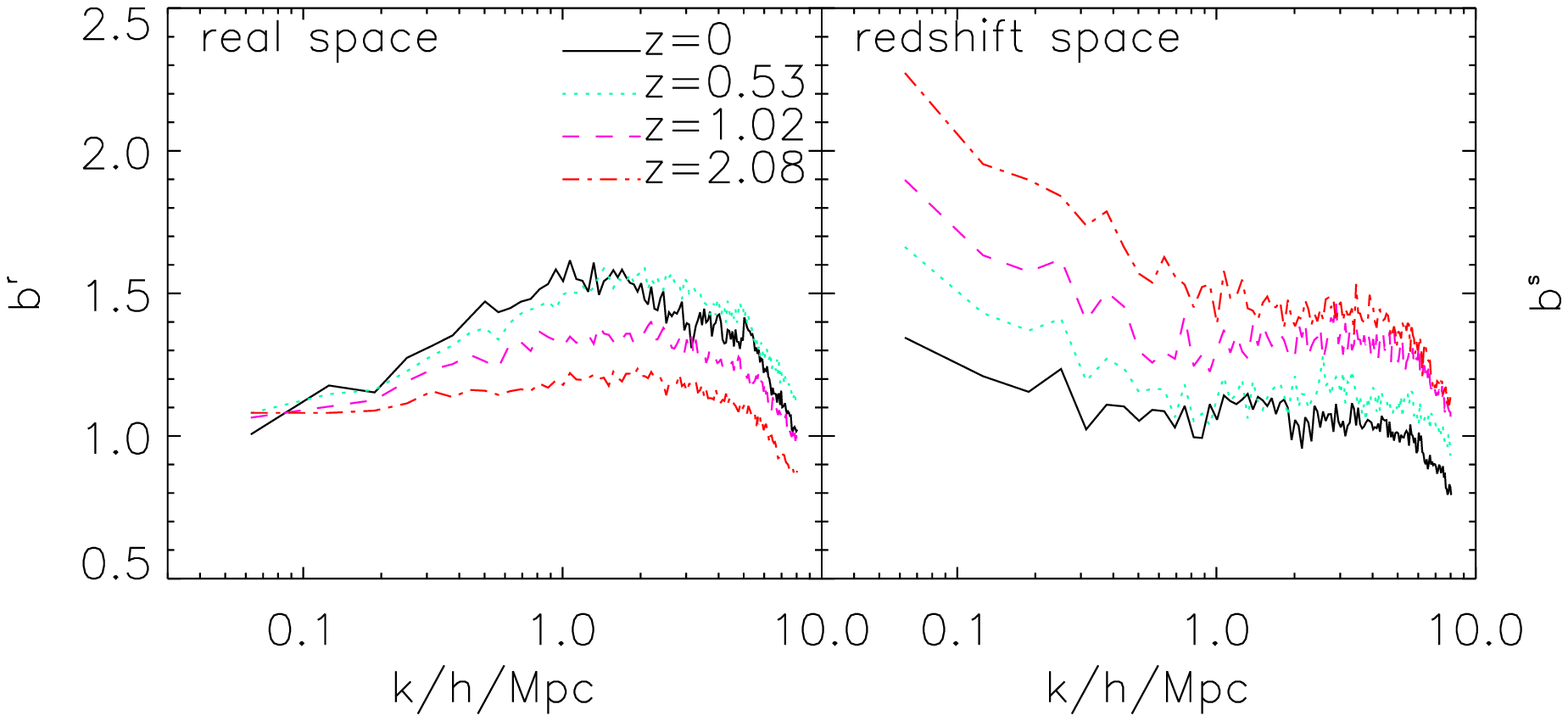}
\caption{The reconstruction bias $b_{\hat{\Theta}}^r$ in the real space (left
panel) and $b_{\hat{\Theta}}^s$ in the redshift space (right panel). In the
real space, the reconstructed signal agree well with $\Theta$ at high
redshifts on large scales up to $k_\perp\le1$\hMpc, while overestimates
$\Theta$ by almost 60\% at recent epochs. However, in the redshift space, the
resulting $\hat{\Theta}^s$ powers are boosted on the largest scales due to the
Kaiser effect, while on smaller scales, FOG effect dominates and brings down
the powers, the severest at z=0.} 
\label{fig:bz}
\efi
As we aim to reconstruct the kSZ map from the galaxy spectroscopic survey,
it's important to assess the reconstruction process. Quantitatively, we use $r$
(Eq. \ref{eqn:r}) and $b_{\hat{\Theta}}$ (Eq. \ref{eqn:b}) to figure out how well
the reconstruction is. As explained in \S \ref{subsec:performance} of the
main context, the S/N of the cross power spectrum (Eq. \ref{eqn:SN}), which is
what we are most concerned about the future observations, is solely dependent on
the determination of the cross correlation coefficient $r$, while a biased
$b_{\hat{\Theta}}$, i.e. the relative amplitude, does not influence the
estimation of the signal to noise of the cross power spectrum. However,
$b_{\hat{\Theta}}$ is a direct measure of the reconstruction bias, and it does
reflect how the reconstructed field suffers from nonlinearity and the redshift
space distortion. So we here address the behavior of $b_{\hat\Theta}$ both in
the real space and the redshift space.

The results of $b^r_{\hat{\Theta}}$ and $b^s_{\hat{\Theta}}$ are shown
respectively in the left and right panel of Fig. \ref{fig:bz}. We
caution the readers to pay attention only to the relative differences between
real and redshift space, or between different redshifts. The absolute value of
$b_{\Theta}$ could be misleading, due to the simplistic assumption of $b_g=1$. 

In real space, $\hat{\Theta}$ tends to more over-estimate $\Theta$ at lower
redshifts ($b_{\Theta}$ increases with redshift).  In redshift space, it is the
opposite case. This is likely caused by the competition between the linear
redshift distortion (the Kaiser effect) and the Finger of God effect. The
Kaiser effect \citep{Kaiser1987} enhances the clustering of matter along the
line of sight and thus the derived  velocity and $\Theta^s$. This is the
reason the redshift space $b_{\Theta}$ at $z=2.08$ is larger than the real
space $b_{\Theta}$ at the same epoch. On the other hand, the Finger of God
effect (see \citealt{Scoccimarro2004} for reviews of the redshift space
distortion and the FOG effect) suppresses the redshift space matter
overdensity and hence $\Theta^s$. The Finger of God effect is amplified by the
nonlinearities with respect to the Kaiser effect, resulting in decreasing
$b_{\Theta}$ with redshift. It is also responsible for the smaller
$b_{\Theta}$ in redshift space at $z=0$ comparing to the one in real space.

\section{The interference of the ``B''-mode velocity}
\label{subsec:app_restimate}
\bfi{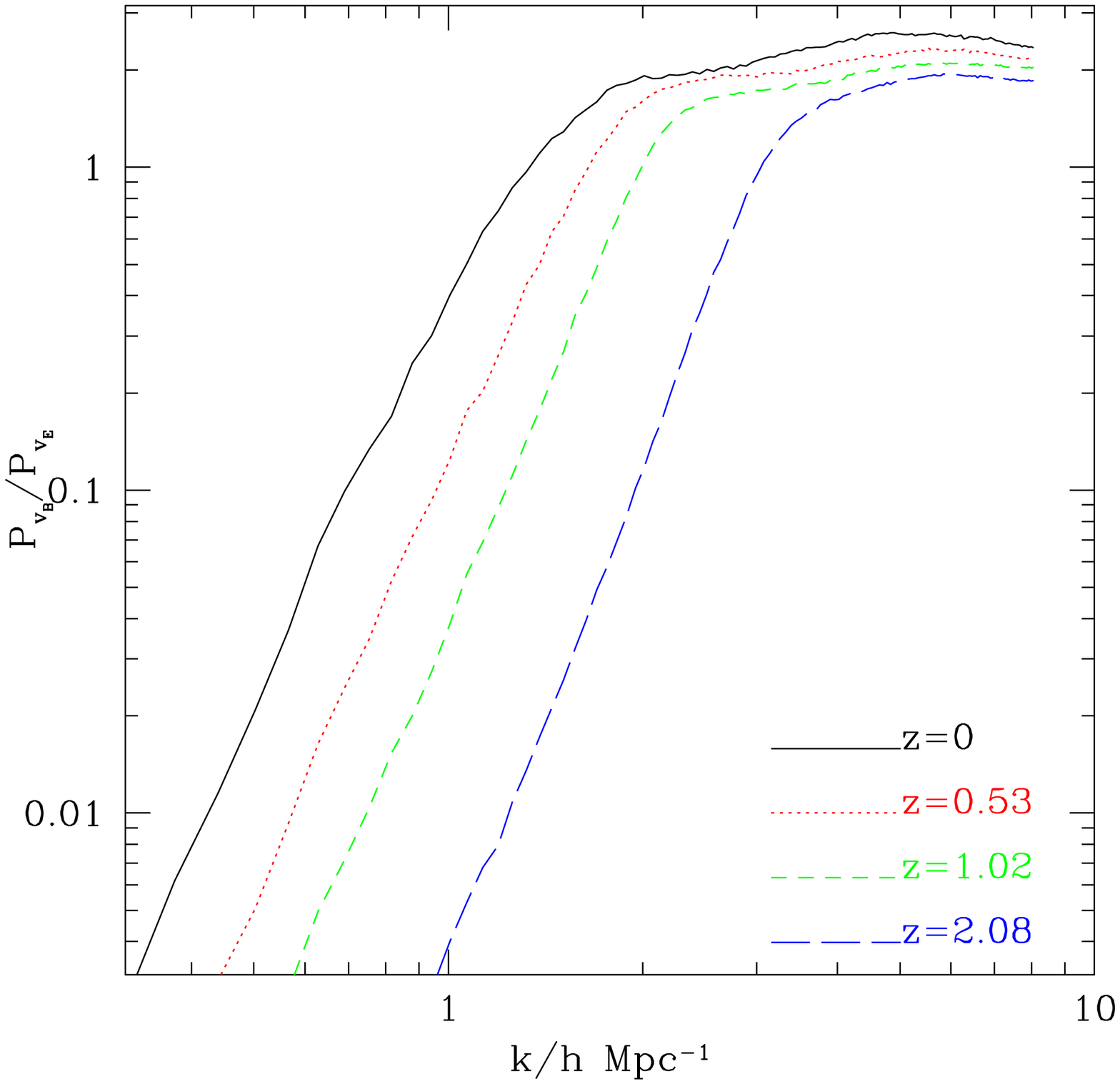}
\caption{The ratio of the power spectra of $E$-mode velocity and $B$-mode
velocity. The results are the ratios of total $B$-mode powers and $E$-mode
powers, different from average method of other figures. ${\bf v}_B$ is
comparable to ${\bf v}_E$ beyond $k\simeq 2$\hMpc at $z=0$, and this
characteristic scale moves to smaller scales at higher redshifts as
velocity is subject to less nonlinearities at early epochs.}
\label{fig:pv3d}
\efi
Our velocity estimator (Eq. \ref{eqn:v}), by construction, can only recover
the irrotational part of the velocity (the so called ``E''-mode velocity,
${\bf v}_E$).  Since ``B''-mode velocity (rotational part of the
velocity, ${\bf v}_B$)  also contributes to the kSZ effect, missing ${\bf
  v}_B$ degrades the reconstruction and the kinetic SZ tomography. This is not
a severe issue at large scales where ${\bf v}_B$ is negligible. However, at
small scales multi-streaming develops and ${\bf v}_B$ begins to
grow. Turbulence in the gas fluid, which can be amplified by gastrophysical
processes like supernovae feedback, also contributes to  ${\bf v}_B$.  To
better understand the impact of ${\bf v}_B$, we directly measure it from our
simulations. 

The $E$-$B$ modes can be obtained straightforwardly in Fourier space by
\be
\begin{split}
    &{\bf v}_E({\bf k})={\bf v}_b({\bf k}) \cdot \hat{\bf k} \ , \\
    &{\bf v}_B({\bf k})={\bf v}_b({\bf k})-{\bf v}_E({\bf k})\ .
\end{split}
\ee
The ratio of the two power spectra in our non-adiabatic run is shown in
Fig. \ref{fig:pv3d}. As  expected, the power spectrum of the ``B''-mode
velocity is much smaller than  the ``E''-mode velocity power spectrum on large
scales. This is especially true at $z\ga 1$ and $k\la 1h/$Mpc. This is one of
the major reasons for the reasonably good performance of the reconstruction at
these scales and redshifts. The B-mode increases towards lower redshifts where
the nonlinearities 
are stronger.  At $z=0$ and $k\ga 1$\hMpc, it is already comparable to the
E-mode. This behavior is largely responsible to  poorer reconstruction at $k\ga
1$\hMpc, especially in redshift space (refer to \S \ref{subsec:scorr}).

\bfi{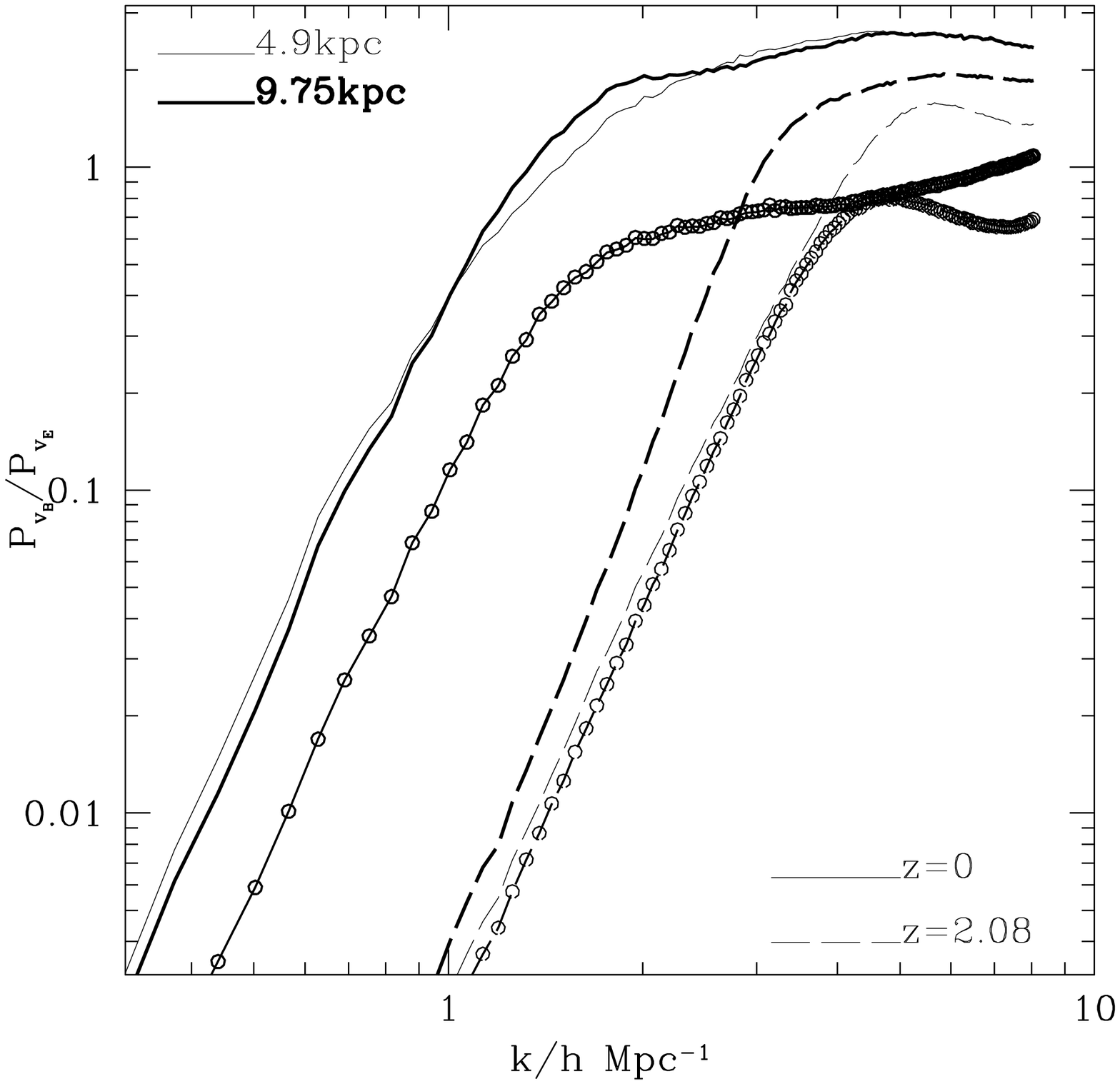}
\caption{The comparison of the gas velocity power spectrum ratio $P_{{\bf
v}_B}/P_{{\bf v}_E}$ in the non-adiabatic simulation (thick lines) and the
adiabatic one (thin lines). Also shown are the ratios of dark matter velocity
power spectrum ratio (lines for adiabatic, open circles for non-adiabatic). The
two simulations have different softening length, with $9.75h^{-1}\,$kpc
(non-adiabatic) vs. $4.9h^{-1}\,$kpc (adiabatic). The difference in softening
length has virtually no impact on  $P_{{\bf v}_B}/P_{{\bf v}_E}$ of dark
matter. So the significant changes in the gas velocity power spectrum between
two simulations is very likely caused by differences in the gastrophysics other than 
difference in the softening length. We need to model ``B''-mode velocity
carefully in the  existence of these gastrophysical processes, an important
ingredient in the kSZ effect modeling.} 
\label{fig:pv}
\efi
Velocity measurement can be tricky in simulations, due to numerical artifacts,
 sampling bias in the velocity assignment and missing gastrophysics
(e.g. \citealt{Pueblas09}).  We do not attempt to perform a detailed study on
these issues. Rather, we present a simple comparison between the two
simulations available.  The two simulations share the same initial conditions
and the same cosmology parameters. However,  the gravitational softening
lengths are different. The one of the non-adiabatic run is $9.75h^{-1}\,$kpc,
twice as large as the adiabatic one, $4.9h^{-1}\,$kpc. With the two
simulations, we can perform interesting observations on the following issues:
\bi
\item Difference between the  velocity field of dark matter and
  that of gas.  Understanding this difference helps improve 
  models on the kSZ effect. 
\item The impact of gastrophysics on the gas velocity.   Radiative and
  dynamical feedback can both cause turbulence in the gas fluid and thus
  affect the B-mode velocity. Comparing the gas velocity in the two
  simulations, we are able to get a handle on this issue. 
\item  The impact of the simulation softening length. The two simulations have
  different softening length. Comparing the dark matter velocity in the two
  simulations, we can estimate its impact.  The reason that we do not use gas
  velocity to do the comparison is that the gas velocity is also affected by
  different gastrophysics in the two simulations. 
\ei

The ratios of two power spectra, $P_{{\bf v}_B}/P_{{\bf v}_E}$, are shown in
Fig. \ref{fig:pv}.  Since dark matter velocity is insensitive to
gastrophysics, differences, if any, between the two simulations are likely
caused by difference in the softening length.  We find the dark matter
velocity has virtually no change in the two simulations. So the influence of
softening length to the  dark matter velocity is negligible. Although we are
not able to directly 
quantify the influence of softening length to the gas velocity, this result
suggests that the influence of softening length should be also under control
for the gas velocity. 

To the opposite, the gas velocity differs significantly between the two
simulations, reflecting the significant influence of gastrophysics. At
$z=2.08$, the B-mode velocity of the non-adiabatic run is a factor of a few
larger than that of the adiabatic run, likely caused by strong star formation
and supernovae feedback at that epoch. Interestingly, this amplification in
the B-mode velocity by gastrophysics at $z=2.08$ is significantly weakened or
even altered at $z=0$. This is likely correlated with the fact that star
formation rate decreases significantly from $z=2$ to $z=0$. 

The vorticity in the gas velocity is larger than that in the dark matter
velocity at all redshifts, all scales and in both simulations. This difference
is likely caused by gas viscosity and dissipation.  It is an important issue
to understand in order to improve the kSZ modeling. 

\end{document}